\documentclass[amsmath,prd,aps,twocolumn,floatfix,nofootinbib]{revtex4-2}

\usepackage{amsfonts}
\usepackage{bm}
\usepackage{graphicx}
\usepackage{color}

\usepackage{graphicx}
\usepackage{dcolumn}
\usepackage{bm}
\usepackage[english]{babel}
\usepackage[utf8]{inputenc}
\usepackage{academicons}
\usepackage{subfigure}
\usepackage{dirtytalk}
\usepackage{amsthm}
\usepackage{mathtools}
\usepackage{physics}
\usepackage{xcolor}
\usepackage{adjustbox}
\usepackage{placeins}
\usepackage[T1]{fontenc}
\usepackage{lipsum}
\usepackage{verbatim}
\usepackage{csquotes}
\usepackage{xspace}
\usepackage[pdftex, pdftitle={Article}, pdfauthor={Author}]{hyperref}
\def\mpchinv{$h^{-1}\mathrm{Mpc}$\xspace}
\def\gpchinv{$h^{-1}\mathrm{Gpc}$\xspace}

\newcommand{\orcid}[1]{\href{https://orcid.org/#1}{\textcolor[HTML]{A6CE39}{\aiOrcid}}}

\newcommand{\utkarsh}[1]{\textcolor{black}{ #1}}

\def\x{{\bf x}}
\def\k{{\bf k}}

\def\fnl{f_{NL}}

\def\nbody{$N$-body\xspace}
\def\npoint{$N$-point\xspace}
\def\s8m{\texttt{s8\_m}\xspace}
\def\s8p{\texttt{s8\_p}\xspace}
\def\bigoh{{\mathcal O}}

\newcommand{\newtopic}[1]{\medskip\par\indent{\bf\em #1}}

\newcommand\nn{\nonumber}
\newcommand\ra{\rightarrow}

\begin{document}

\title{Constraining $f_{NL}$ using the large-scale modulation of small-scale statistics}

\author{Utkarsh Giri}
    \email[Correspondence email address: ]{utkarshgiri18@gmail.com}
    \affiliation{Department of Physics, University of Wisconsin-Madison, Madison, WI 53706, USA}
    
\author{Moritz M\"unchmeyer}
    \affiliation{Department of Physics, University of Wisconsin-Madison, Madison, WI 53706, USA}

\author{Kendrick M. Smith}
    \affiliation{Perimeter Institute for Theoretical Physics, Waterloo, ON N2L 3G1, Canada}

\date{\today}

\begin{abstract}

We implement a novel formalism to constrain primordial non-Gaussianity of the local type from the large-scale modulation of the small-scale power spectrum. Our approach combines information about primordial non-Gaussianity contained in the squeezed bispectrum and the collapsed trispectrum of large-scale structure together in a computationally amenable and consistent way, while avoiding the need to model complicated covariances of higher $N$-point functions. This work generalizes our recent work, which used a neural network estimate of local power, to the more conventional local power spectrum statistics, and explores using both matter field and halo catalogues from the Quijote simulations. 
We find that higher $N$-point functions of the matter field can provide strong constraints on $\fnl$, but higher $N$-point functions of the halo field, at the halo density of Quijote, only marginally improve constraints from the two-point function.
\end{abstract}

\maketitle

\section{Introduction} 
\label{sec:introduction}

A main target of upcoming galaxy surveys of large-scale structure (LSS) like \textsc{DESI} \cite{DESI:2016fyo}, \textsc{SPHEREx} \cite{dore2014cosmology}, \textsc{Euclid} \cite{EUCLID:2011zbd} and \textsc{Rubin Observatory} \cite{LSSTScience:2009jmu} is to detect and characterize any non-Gaussianity in the primordial fields. Different models of inflation, as well as alternatives to inflation, predict primordial non-Gaussianity of various kinds which are sensitive to the field and energy content of the early universe \cite{chen2010primordial, Achucarro:2022qrl}. Among shapes of non-Gaussianity, the so called local type parameterized by $f_{NL}$\footnote{Throughout this work $\fnl$ refers to local type primordial non-Gaussianity. We prefer it over the more common $\fnl^{loc}$ parameterization for simplicity and conciseness.}, which can detect multi-field inflation, is the most experimentally accessible with upcoming galaxy surveys. The primordial potential $\Phi$ under this parameterization is given by \cite{Komatsu:2001rj}
\begin{equation}
     \Phi(\x) = \Phi_G(\x) + f_{NL}(\Phi_G(\x)^2 - \langle\Phi_{G} ^2\rangle) \label{phi_ng}
\end{equation}
where $\Phi_G$ is an auxiliary Gaussian field, and the phenomenological parameter $f_{NL}$ controls the level of non-Gaussianity. Cosmic Microwave Background (CMB) experiments like Planck \cite{Planck:2018vyg} have put strong constraints on $\fnl$ using the bispectrum statistics of temperature and polarization maps. Further improvement is expected from LSS surveys in the coming years \cite{Alvarez:2014vva}. Although the lowest-order \npoint statistic sensitive to $\fnl$ in a weakly non-Gaussian universe is the bispectrum, the details of structure formation in an $\fnl \ne 0$ universe lead to \emph{scale-dependent bias} in the power spectrum of density tracers like halos on the largest scales \cite{Dalal:2007cu, Slosar_2008, dePutter:2016moa, Biagetti_2019}, making halo bias a promising and clean probe of $\fnl$ in LSS. Its constraining power is expected to surpass the Planck CMB constraints in the very near future \cite{Sailer:2021yzm}. In an $\fnl \ne 0$ universe, the halo power spectrum acquires a characteristic $1/k^2$ scaling that cannot be mimicked by any other physical process. However, the halo power spectrum is not the only statistic sensitive to $\fnl$, and attempts to develop new statistics are ongoing. Several recent works have tried to quantify the information gains one can hope to achieve from probing smaller scales. Approaches like field-level modelling \cite{Andrews:2022nvv}, one-point analysis \cite{Friedrich2019}, and topological analysis \cite{Biagetti:2020skr}, among others, have shown promise in extracting more information. In \cite{giri_2023}, we developed a neural network enhanced approach for constraining $\fnl$ and obtained significantly better constraints on $\fnl$ using information from the small-scale density field, while preserving the robustness of the scale-dependent bias analysis. 

In this work, we go back to a more traditional approach and ask the question ``how much $f_{NL}$ information does the halo bispectrum add to the halo power spectrum?''.
A number of recent papers have attempted to partially or completely answer this question \cite{Jeong_2009,Baldauf_2011,Schmittfull_2015,Chiang_2017,deputter2018primordial,Dizgah_2020,Dai_2020,darwish2020density,MoradinezhadDizgah:2020whw}.

There are several (related) challenges which must be addressed to answer this question.
First, enumerating all nuisance parameters which must be marginalized is nontrivial, especially on quasilinear and smaller scales.
One approach is the Effective Theory of LSS \cite{Cabass:2022ymb, DAmico:2022gki}.
Next, the observables (i.e.\ 2-point and 3-point functions) must be modelled in enough detail to marginalize nuisance parameters.
Finally, it has been shown \cite{deputter2018primordial,Barreira:2019icq,Floss:2022wkq} that modelling the covariance between observables is critical.
This is particularly challenging, since the covariance of (2+3)-point functions involves (4,5,6)-point functions which are difficult to compute.

Before explaining our approach, we will highlight two recent studies which are closely related to this paper.
First, the Quijote-PNG collaboration \cite{qpng1, jung2022quijotepng, coulton2022quijote, Jung:2022gfa} recently studied the $\fnl$ information content of $N$-body simulations, by running enough simulations that the (2+3)-point observables and their covariance could be directly estimated from Monte Carlo simulations.
This approach is very flexible, since it incorporates bispectrum information with arbitrary $(k_1,k_2,k_3)$ dependence, and applies to any form of primordial non-Gaussianity, not just the local type considered in this paper.
However, $N$-body simulations are computationally expensive, especially when pushed to scales where baryonic feedback is important \cite{Angulo_2021,chisari2019modelling, Villaescusa-Navarro:2020rxg}.
Moreover, in a purely simulation-based approach it is difficult to marginalize a conservative set of nuisance parameters (such as effective field theory (EFT) parameters). 
The Quijote-PNG analysis marginalizes $\Lambda$CDM cosmological parameters, and a minimum halo mass $M_{\rm min}$.
With these caveats, the main result of \cite{qpng1, jung2022quijotepng, coulton2022quijote, Jung:2022gfa} is that the 3-point function of the {\em matter} field contains significant $\fnl$ information, but the 3-point function of the halo field does not (in the sense that $\sigma(\fnl)$ in a (2+3)-point analysis is marginally better than a 2-point analysis).

Second, Goldstein et al \cite{Goldstein:2022hgr} studied $\fnl$ information in the matter bispectrum.
The key idea of this paper is to restrict attention to the squeezed bispectrum\footnote{Recall that a three-point function (or bispectrum)
    \begin{equation}
    \big\langle \delta({\bf k_L}) \delta({\bf k_S}) 
      \delta({\bf k_S}')\big\rangle \nonumber
    \end{equation}
    is said to be {\em squeezed} if $k_L \ll \min(k_S,k_S')$.
    A four-point (or trispectrum) configuration
    \begin{equation}
    \big\langle \delta({\bf k_1}) \delta({\bf k_2}) 
      \delta({\bf k_3})  \delta({\bf k}_4) \big\rangle
      \nonumber
    \end{equation}
    is said to be {\em collapsed} if $|{\bf k}_1 + {\bf k}_2| \ll \min(k_1,k_2,k_3,k_4)$. }, rather than using the full bispectrum. 
Then there are consistency relations \cite{Peloso, Kehagias, Simonovic, Esposito:2019jkb} which show that in the squeezed limit $k_L \ll k_S$, the leading term in the bispectrum is proportional to $(\fnl/k_L^2)$.
On the other hand, non-primordial physics produces contribution with a ``softer'' scale dependence $\bigoh(k_L^0) + \bigoh(k_L^2) + \cdots$.
By marginalizing a general contribution of this softer type, one automatically marginalizes all physical nuisance parameters (cosmological or astrophysical).

Our approach is similar in spirit to \cite{Goldstein:2022hgr}, but extended as follows. As shown in \cite{Smith:2007rg, Smith:2011if, Chiang:2014oga}, the squeezed bispectrum can be equivalently represented as a cross power spectrum $P_{\delta_m \pi}$, where the field $\pi(\x)$ is the locally measured small-scale power spectrum (see Eq.\ \ref{eq:pmm_loc_def} below for precise definition). Similarly, the collapsed trispectrum can be represented as the auto power spectrum $P_{\pi\pi}$ \cite{Smith:2011if, deputter2018primordial}.

We then argue heuristically (and verify with Quijote-PNG simulations) that on large scales, the field $\pi(\x)$ is described by a linear bias model of schematic form (see Eq.\ \ref{eq:key_conjecture1} for precise form):
\begin{equation}
\pi(\k) \sim \left( b_\pi + \beta_\pi \frac{f_{NL}}{k^2} \right) \delta_m(\k) + \big(\mbox{white noise} \big) \label{eq:intro_bias_model}
\end{equation}
Using this bias model, it is straightforward to write down a ``field-level'' likelihood function using large-scale modes of $\delta_m$ and $\pi$ (Eq.\ \ref{eq:mode_likelihood} below), which captures $\fnl$ information from the squeezed bispectrum ($P_{m\pi}$) and collapsed trispectrum ($P_{\pi\pi}$).
When we sample the likelihood, we marginalize all Gaussian bias parameters $b_\pi$ (but not non-Gaussian biases $\beta_\pi$), and a set of parameters $N_{\pi\pi'}$ describing the white noise.
We retain sensitivity to $\fnl$ because the non-Gaussian signal in Eq.\ (\ref{eq:intro_bias_model}) has a characteristic $(1/k^2)$ scale dependence (like non-Gaussian halo bias or the squeezed bispectrum from \cite{Goldstein:2022hgr}) which makes the analysis robust to uncertain cosmological and astrophysical parameters, without needing to enumerate these parameters explicitly.

The simplicity and low computational cost of our approach makes it easy for us to explore variants of the analysis -- for example we study bispectra and trispectra of the halo field $\delta_h$ (rather than the matter field $\delta_m$) in \S\ref{halo_based_analysis}.
The basic idea of encoding higher-point information in the large-scale modes of an auxiliary $\pi$-field originated in \cite{giri_2023}, where we constructed a $\pi$-field using a convolutional neural network. A CNN can learn a $\pi$-field that gives optimal $\fnl$ constraints and can thus improve over the locally measured small-scale power spectrum which we use here, at the cost of introducing a machine learning element.

We present our approach and develop an end-to-end mode based MCMC pipeline to explore the constraining power of the formalism.  We analyze the matter field from the {Quijote} \nbody simulation datasets  \cite{Villaescusa-Navarro:2019bje} focusing particularly on the non-linear regime. We then analyze halo catalogues from the corresponding set of simulations. After presenting our formalism in \ref{sec:formalism} and \ref{sec:fnlcosmo}, in section \ref{sec:simulations}, we present details of the {Quijote} simulation suite and our processing pipeline, which we use for validation and analysis. In section \ref{sec:results}, we present the results from our MCMC analysis. Finally in \ref{sec:discussion}, we present our conclusions.

\section{Formalism}
\label{sec:formalism}

In this section we will describe our approach intuitively with details postponed to subsequent sections. For conciseness, we present the formalism in terms of the halo overdensity field $\delta_h$, but subsequent sections will generalize this to other cosmological fields (like the matter field $\delta_m$).

Suppose we are interested in constraining $f_{NL}$ using the squeezed halo bispectrum
\begin{equation}
\big\langle \delta_h(k_L) \delta_h(k_S) \delta_h(k_S') \big\rangle 
\hspace{1cm} \mbox{where } k_L \ll k_S  \label{eq:squeezed_delta_h}
\end{equation}
and suppose we further assume that we average over a wide bin in $k_S$.

The squeezed bispectrum (\ref{eq:squeezed_delta_h}) can be viewed more intuitively as the cross power spectrum $P_{\delta_h \pi}(k_L)$, where $\pi(\k_L)$ is the locally observed small-scale halo power spectrum, integrated over a wide range of wavenumbers $k_S$.
(For the formal definition of ``locally observed small-scale power spectrum'', see \S\ref{ssec:observables} -- here we just note that $\pi$ is built quadratically out of small-scale modes of $\delta_h$.)

On large scales, the halo overdensity can famously be modelled as (schematically) 
\begin{equation}
\delta_h(\k_L) = \left( b_h + \beta_h \frac{f_{NL}}{k_L^2} \right) \delta_m(\k_L) 
 + \epsilon_h(\k_L)
\label{eq:intro_delta_h}
\end{equation}
where $b_h$ and $\beta_h$ are the usual Gaussian and non-Gaussian halo bias parameters respectively, while $\epsilon_h(\k_L)$ is the Poisson noise.
We will argue that the new field $\pi(\k_L)$ can be modelled on large scales using a similar linear bias model:
\begin{equation}
\pi(\k_L) = \left( b_\pi + \beta_\pi \frac{\fnl}{k_L^2} \right) \delta_m(\k_L)
 + \big[ \mbox{Noise } \epsilon_\pi(\k_L) \big]
\label{eq:intro_pi}
\end{equation}
where the power spectra of the noise fields $\epsilon_h$, $\epsilon_\pi$ approach constants as $k \rightarrow 0$:
\begin{equation}
\left( \begin{array}{cc}
  P_{\epsilon_h\epsilon_h}(k) & P_{\epsilon_h\epsilon_\pi}(k) \\
  P_{\epsilon_h\epsilon_\pi}(k) &  P_{\epsilon_\pi\epsilon_\pi}(k)
\end{array} \right) \rightarrow 
\left( \begin{array}{cc}
  N_{hh} & N_{h\pi} \\
  N_{h\pi} & N_{\pi\pi}
\end{array} \right)
\hspace{0cm} \mbox{as } k\ra 0
\label{eq:intro_noise}
\end{equation}

These expressions involve some new parameters $b_\pi$, $\beta_\pi$, $N_{h\pi}$, and $N_{\pi\pi}$ (Note that $N_{hh} \approx 1/n_h$, where $n_h$ is the halo number density).
These new parameters would be difficult to calculate analytically (e.g. in the halo model) but we will describe a brute force procedure for estimating them from an ensemble of $N$-body simulations, in \S\ref{ssec:estimating_bias_and_noise}.

After all parameters in Eqs.~(\ref{eq:intro_delta_h})--(\ref{eq:intro_noise}) have been determined, we have a simple picture with two fields $\delta_h,\pi$ with $\fnl$-dependent power spectra.
The original question, ``how much $\fnl$ information does the halo bispectrum add to the halo power spectrum?'' can be rephrased as the question ``how much $\fnl$ information does $P_{h\pi}$ add to $P_{hh}$?''.

We can also consider several generalizations of the approach as follows:
\begin{itemize}
    \item In the two-field picture, one could also ask how much $\fnl$ information can be obtained if $P_{\pi\pi}$ is included (in addition to $P_{hh}, P_{h\pi}$).
    In $N$-point language, this corresponds to asking how much $\fnl$ information is obtained if the {\em collapsed}
     halo trispectrum is included (in addition to the halo power spectrum and squeezed bispectrum).
    \item So far, we have assumed for simplicity that the squeezed bispectrum is integrated over a wide range of small-scale wavenumbers $k_S$. This implicitly assumes a fixed $k_S$-weighting. To allow an arbitrary weighting, we could define a few $k_S$-bins, and define one $\pi$ field for each bin. Then, instead of having two large-scale fields $(h,\pi)$, we would have $(N+1)$ fields $(h,\pi_i)$, where $N$ is the number of $k_S$-bins. Similarly, we could allow an arbitrary halo mass weighting by defining multiple $h$ and $\pi$ fields corresponding to different halo mass bins.

    \item Instead of using the halo field $\delta_h$, one could assume that the matter field $\delta_m$ can be observed on small scales, to study the $\fnl$ information limit for the true, noiseless matter field. (We will do this in \S\ref{sec:matter_based_analysis}, before moving on to the halo case in \S\ref{halo_based_analysis}.)
    
    \item We could replace the locally measured small-scale power spectrum $\pi$ (a quadratic function of $\delta_h$) by a more complex nonlinear function. For example, one could use a convolutional neural network (CNN), or the wavelet scattering transform (WST) \cite{Cheng__2020}.
    In \cite{giri_2023}, we constructed a nonlinear field $\pi(\x)$ from small-scale modes of the matter field, using a CNN that was optimized for sensitivity to $\fnl$.
    In \cite{Sullivan:2023qjr}, an $\fnl$-optimized observable was constructed from galaxy catalogs (including color information) using machine learning methods.

\end{itemize}

In the following sections, we describe our formalism in detail.
\subsection{Fourier conventions}
\label{ssec:fourier_conventions}

Our Fourier conventions in a finite pixelized box, with box volume $V_{\rm box}$ and pixel volume $V_{\rm pix}$, are:
\begin{align}
\phi(\x) &= V_{\rm box}^{-1} \sum_{\k} \phi(\k) e^{i\k\cdot\x} \\
\phi(\k) &= V_{\rm pix} \sum_{\x} \phi(\x) e^{-i\k\cdot\x} \\
\big\langle \phi(\k) \psi(\k')^* \big\rangle &= V_{\rm box} P_{\phi\psi}(k) \, \delta_{\k\k'}
\end{align}

\subsection{Observables}
\label{ssec:observables}

We define an {\em observable} of an $N$-body simulation to be a 3D field $\pi(\x)$ which is derived from the simulation, in a way which preserves the symmetries of the simulation volume (translations and permutations/reflections of the axes). Here are some examples of observables:
\begin{itemize}
    \item The matter density field $\rho_m(\x)$.
    \item The halo number density field $n_h(\x)$, for some choice of halo mass bin (or halo mass weighting).
    \item The locally measured small-scale matter power spectrum
    \begin{equation}
        P_{mm}^{\rm loc}(\x) = \left( \int \frac{d^3\k}{(2\pi)^3} \, W(k) \rho_m(\k) e^{i\k\cdot\x} \right)^2
        \label{eq:pmm_loc_def}
    \end{equation}
    where $W(k)$ is a high-pass filter peaked at some characteristic small scale $k_S$. (The normalization of $P_{mm}^{\rm loc}$ in Eq.~(\ref{eq:pmm_loc_def}) is arbitrary.)
    \item Similarly, given a choice of halo mass bin (or halo mass weighting), we can define the locally measured small-scale halo power spectrum
        \begin{equation}
        P_{hh}^{\rm loc}(\x) = \left( \int \frac{d^3\k}{(2\pi)^3} \, W(k) n_h(\k) e^{i\k\cdot\x} \right)^2
        \label{eq:phh_loc_def}
    \end{equation}
\end{itemize}

\subsection{Bias and noise}

We now provide rigorous definitions for some central quantities used throughout this paper. In particular, we define the Gaussian bias (${b_\pi}$), the non-Gaussian bias ($\beta_{\pi}$) as well as the noise $N_{\pi\pi}$ for an observable $\pi$ described in the last section. %

The large-scale (Gaussian) {\em bias} $b_\pi$ is defined by:
\begin{equation}
\lim_{k\ra 0} P_{\pi \delta_m}(k) = b_\pi P_{mm}(k) \label{gaussian_bias_def}
\end{equation}
Another way of thinking about bias is:
\begin{equation}
\pi(\k) = b_\pi \delta_m(\k) + (\mbox{Noise field $\epsilon_\pi(\k)$})
\end{equation}
where the noise field $\epsilon_\pi$ defined by this equation is uncorrelated with $\delta_m$ on large scales.
In the special case where $\pi = n_h$ is the halo number density field, then $b_\pi = \bar n_h b_h$, where $b_h$ is the usual halo bias.

For a pair of observables $\pi, \pi'$, we define the {\em noise} $N_{\pi\pi'}$ by:
\begin{equation}
\lim_{k\ra 0} P_{\pi\pi'}(k) = b_\pi b_{\pi'} P_{mm}(k) + N_{\pi\pi'}
\end{equation}
If observables $\pi_1, \cdots, \pi_n$ are halo number density fields $\pi_i = n_{h_i}$ corresponding to different mass bins, then $N_{\pi_i\pi_j} = \bar n_{h_i} \delta_{ij}$. This statement assumes a Poisson noise model for halos.

Finally, we define the {\em non-Gaussian bias} $\beta_\pi$ by:\footnote{When we write $(\partial/\partial\log\sigma_8)$, we really mean a derivative $(\partial/\partial\log\Delta_\zeta)$ with respect to the overall amplitude of the initial adiabatic curvature power spectrum $P_\zeta(k) = 2\pi\Delta_\zeta^2 (k/k_{\rm piv})^{n_s-4}$. %
}
\begin{equation}
\beta_\pi \equiv \frac{\partial\bar\pi}{\partial\log\sigma_8}
\label{eq:beta_def}
\end{equation}
where the quantity $\bar\pi$ is the mean value of $\pi(\x)$, taken over both Monte Carlo simulations and spatial pixels.

If $\pi = n_h$ is the halo number density field, then the non-Gaussian bias $\beta_\pi$ is given by the famous equation (with an extra factor $\bar n_h$ since we use $\pi=n_h$ instead of $\pi=\delta_h$):
\begin{equation}
\beta_\pi \approx \bar n_h \delta_c (b_g - 1)  \label{eq:beta_pi_nh}
\end{equation}
where $\delta_c \approx 1.42$.
This is really an approximation to the true non-Gaussian bias $\beta_h \equiv (\partial\bar n_h/\partial\log\sigma_8)$. The approximation in Eq.~(\ref{eq:beta_pi_nh}) is motivated by spherical collapse models of halo formation, and is usually accurate to $\sim$10\% when compared with simulations \cite{Desjacques_2008, Biagetti_2017}.

The non-Gaussian bias $\beta_\pi$ parametrizes the level of excess clustering on large scales in an $\fnl$ cosmology, in a sense that we will make precise in the next section.

\section{$f_{NL}$ cosmology}
\label{sec:fnlcosmo}

\subsection{Key conjectures}
\label{ssec:key_conjectures}

In the previous section, we defined the bias $b_\pi$ and noise $N_{\pi\pi'}$ by the the large-scale ($k\ra 0$) power spectra:
\begin{align}
P_{g\pi}(k) &\rightarrow b_\pi P_{mm}(k)  \label{eq:key_conj_pre1} \\
P_{\pi\pi'}(k) & \rightarrow b_\pi b_{\pi'} P_{mm}(k) + N_{\pi\pi'} \label{eq:key_conj_pre2}
\end{align}
So far, we have assumed $f_{NL}=0$.
In this section, we will generalize Eqs.\ (\ref{eq:key_conj_pre1}), (\ref{eq:key_conj_pre2}) to an $\fnl$ cosmology.
We will describe these results as ``conjectures'', to emphasize that they are predictions that we will verify with simulations later 
 (\S\ref{sec:simulations}).

\newtopic{Key conjecture 1.}
In an $f_{NL}$ cosmology, Eq.~(\ref{eq:key_conj_pre1}) generalizes as (on large scales):
\begin{equation}
P_{m\pi}(k) = \left( b_\pi + 2 \beta_\pi \frac{f_{NL}}{\alpha(k,z)} \right) P_{mm}(k)
\label{eq:key_conjecture1}
\end{equation}
where $\beta_\pi$ was defined in Eq.\ (\ref{eq:beta_def}). The function $\alpha(k,z)$ is defined by:
\begin{equation}
\alpha(k,z) \equiv \frac{2k^2 T(k) D(z)}{3\Omega_m H_0^2}
\label{eq:alpha_def}
\end{equation}
so that $\delta_m(k,z) = \alpha(k,z) \Phi(k)$, where $\Phi(k)$ is the primordial potential from Eq.\ \ref{phi_ng}.

Note that $\alpha(k,z) \propto k^2$ as $k\rightarrow 0$, so the key conjecture (\ref{eq:key_conjecture1}) predicts that any observable with $\beta_\pi \ne 0$ has large-scale bias proportional to $(f_{NL}/k^2)$. This generalizes the famous non-Gaussian halo bias in the case $\pi = \rho_h$.

\newtopic{Key conjecture 2.}
In an $f_{NL}$ cosmology, Eq.~(\ref{eq:key_conj_pre2}) generalizes as (on large scales):
\begin{align}
P_{\pi\pi'}(k) &= \left( b_\pi + 2 \beta_\pi \frac{f_{NL}}{\alpha(k,z)} \right)  \nonumber \\
&\times \left( b_{\pi'} + 2 \beta_{\pi'} \frac{f_{NL}}{\alpha(k,z)} \right)
P_{mm}(k) + N_{\pi\pi'}
\label{eq:key_conjecture2}
\end{align}
That is, the large-scale cross spectrum $P_{\pi\pi'}(k)$ has the ``minimal'' form expected from the bias model (\ref{eq:key_conjecture1}), plus a white noise term $N_{\pi\pi'}$.%

\newtopic{Key conjecture 3.}
Given $N$ observables $\pi_1,\cdots,\pi_N$, the matter field $\delta_m(\k)$ and the observables $\pi_i(\k)$ are Gaussian fields for sufficiently small $k$. Thus, the higher point statistics and likelihood function of the field realizations are determined by the power spectra in key conjectures 1 and 2.

\subsection{Schematic derivation of key conjectures 1 and 2}
\label{sec:key_conjecture_derivation}

\noindent In an $\fnl$ cosmology, the initial conditions are given by:
\begin{equation}
\Phi(\x) = \Phi_G(\x) + \fnl(\Phi_G(\x)^2 - \langle \Phi_G^2 \rangle)  \label{eq:pbs_fnl}
\end{equation}
To analyze the effect of a long-wavelength mode, let us 
decompose the {\em Gaussian} potential as a sum $\Phi_G = \Phi_l + \Phi_s$ of long-wavelength and short-wavelength 
contributions. The long/short-wavelength decomposition of the non-Gaussian potential $\Phi$ is then
\begin{align}
\Phi(\x) &= \underbrace{ \Phi_l(\x) + \fnl \left(\Phi_l(\x)^2 - \langle \Phi_l^2 \rangle \right) }_{\rm long} \nonumber \\
  &+ \underbrace{(1 + 2\fnl\Phi_l(\x)) \Phi_s(\x) + \fnl( \Phi_s(\x)^2 - \langle \Phi_s^2 \rangle )}_{\rm short}
  \label{eq:longshort}
\end{align}
and contains explicit coupling between long and short wavelength modes of the Gaussian potential. 

The term $(1 + 2\fnl\Phi_l(\x)) \Phi_s(\x)$ in Eq.~(\ref{eq:longshort}) may be interpreted as follows.
In a local region where the long-wavelength potential takes some value $\Phi_l$, the overall amplitude of the small-scale
modes is multiplied by a factor $(1 + 2\fnl\Phi_l)$.
That is, the ``locally observed'' value of $\sigma_8$ fluctuates throughout the universe, and is given on large scales by:
\begin{equation}
\sigma_8^{\rm loc}(\x) = \big( 1 + 2 \fnl \Phi_l(\x) \big) \, \bar\sigma_8 
\label{eq:pbs_sigma8_loc}
\end{equation}
These large-scale variations in $\sigma_8$ induce large-scale variations in the observable $\pi$ as follows:
\begin{equation}
\pi(\x) = 
b_\pi \delta_m(\x)
+ \beta_\pi \log\left( \frac{\sigma_8^{\rm loc}(\x)}{\bar\sigma_8} \right)+\big(\mbox{uncorrelated noise} \big)
\label{eq:pbs_pi0}
\end{equation}
Here, the first and third terms arise in an $\fnl=0$ cosmology.
The second term is new, and arises because a fluctuation $(\delta \log\sigma_8^{\rm loc})$ of sufficiently long wavelength has the same effect on the observable $\pi$ as a shift $(\delta\log\sigma_8)$ in the ``background'' cosmological parameter $\sigma_8$.

Combining Eqs.\ (\ref{eq:pbs_sigma8_loc}),~(\ref{eq:pbs_pi0}) and writing $\delta_m(k,z) = \alpha(k,z) \Phi(k)$, we obtain the following expression for $\pi(\x)$ on large scales:
\begin{equation}
\pi(\k) = \left( b_\pi + 2\beta_\pi \frac{f_{NL}}{\alpha(k,z)} \right) \delta_m(\k) + \big(\mbox{uncorrelated noise} \big)
\label{eq:pbs_pi}
\end{equation}
If we cross-correlate Eq.~(\ref{eq:pbs_pi}) with $\delta_m$, the noise term goes away, and we get:
\begin{equation}
P_{m\pi}(k) = \left( b_\pi + 2\beta_\pi \frac{f_{NL}}{\alpha(k,z)} \right) P_{mm}(k)
\end{equation}
which is key conjecture 1.
If we cross-correlate the expression for $\pi(\k)$ in Eq.~(\ref{eq:pbs_pi}) with a similar expression for a different observable $\pi'(\k)$, we get:
\begin{align}
P_{\pi\pi'}(k) &= \left( b_\pi + 2 \beta_\pi \frac{f_{NL}}{\alpha(k,z)} \right) \nonumber \\
 &\times \left( b_{\pi'} + 2 \beta_{\pi'} \frac{f_{NL}}{\alpha(k,z)} \right)
P_{mm}(k) + N_{\pi\pi'}
\end{align}
which is key conjecture 2. We validate both these conjectures using \nbody simulations in the next section. In the linear regime, the noise $N_{\pi\pi'}$ would be inversely proportional to the number of local modes per unit volume, while in the non-linear regime we use here, it is a more complicated function.

\section{Simulation pipeline}
\label{sec:simulations}

\subsection{Simulations}

A major challenge in  simulation-based studies for constraining $\fnl$ is the need for large sets of large-scale cosmological simulations with sufficiently high resolution, ideally run with both Gaussian as well as non-Gaussian initial conditions. Several collaborative efforts have been made recently to release massive suites of simulations, including both hydrodynamical and dark matter only simulations, for broad public use. In this paper, we utilize the Quijote suite of simulations \cite{Villaescusa-Navarro:2019bje}, which consists of 44,100 publicly accessible, full \nbody simulations that cover over 7,000 cosmological models within the cosmological parameter hyperplane, with varying particle resolution.  These simulations are run using the {Gadget-III} \cite{Springel:2005mi} simulation code.

One of the primary aims of the Quijote simulations is to quantify information content on cosmological observables and as such, there are sets of simulations where a single parameter is perturbed above or below its fiducial value to facilitate numerical derivative computations using finite difference. In our work, we will be using the \texttt{fiducial} simulations to test our formalism along with the datasets \texttt{s8\_m} and \s8p which we use to estimate non-Gaussian bias $\beta$ (see Eq.\ (\ref{eq:piestimator}) below). The \texttt{fiducial} simulations use $\Omega_{\rm m}=0.3175$, $\Omega_{\rm b}=0.049$, $h=0.6711$, $n_s=0.9624$, $w=-1$ and $\sigma_8=0.834$ and are run in a box of size 1 (\gpchinv)$^3$ using 512$^3$ dark-matter particles to sample the matter field. The initial particle positions and velocities are generated using second-order Lagrangian perturbation theory \cite{Crocce_2006} at redshift $z=127$ and with a Gaussian primordial potential with an effective $\fnl=0$. For the \texttt{s8\_m} simulations, the $\sigma_8$ is lowered to 0.819 while \texttt{s8\_p} simulations have $\sigma_8=0.849$. For all these simulations, the halo catalogue is generated using the classical Friends-of-Friend (FoF) algorithm \cite{fof} with a minimum particle requirement of 20 and a linking length of 0.2. 
To test our formalism for $\fnl$ universe, we make use of Quijote-PNG simulations \cite{coulton2022quijote}. The Quijote-PNG are another large suite of \nbody simulations which extend the Quijote simulations to include simulations with different types of primordial non-Gaussianity. They are run with PNG of local, equilateral and orthogonal type, each characterized by the parameter $\fnl$, with other cosmological parameters and simulation specifications kept identical to the fiducial Quijote simulations. There are two subsets, \texttt{LC\_p} and \texttt{LC\_m} which have local PNG with $\fnl$ set to 100 and -100 respectively. In this study we make use of the \texttt{LC\_p} suite.

\subsection{Generating the $\pi$ fields from density fields}
\label{sec:pi_definition}
For our study with Quijote simulations, we choose to work with simulations  at $z=0$. We use the public library \texttt{Pylians3} \cite{Pylians} to read binary snapshot files and use the cloud-in-cell algorithm \cite{hockneyeastwood} implemented in \texttt{nbodykit} \cite{Hand:2017pqn} to paint particle positions on a 3D mesh to obtain the matter field $\rho_m$. We similarly paint the positions of the halos to obtain the halo number density field $\rho_h$. For both matter and halo field, our 3D mesh grid is of size $1024^3$. This choice of mesh size corresponds to Nyquist frequency 3.2 $h$ Mpc$^{-1}$. This allows us to tap information from the deeply non-linear regime.
The process of painting particle positions on a 3D mesh is known to suffer from an effect called ``aliasing'' which can potentially contaminate modes near the Nyquist frequency. Sampling and resolution related issues also become pronounced on small scales. However, we will show in \S\ref{sec:validation} that our key conjectures from \S\ref{ssec:key_conjectures} are still valid if we construct $\pi$-fields from wavenumbers near the Nyquist frequency. 

We now define 5 matter-derived fields $\{ \pi^{m}_i \}_{i=1,2,3,4,5}$, and 2 halo-derived fields $\{ \pi^{h}_i \}_{i=1,2}$, which will be used extensively throughout the paper.
Each such $\pi$-field is defined by:
\begin{equation}
    \pi^{f}_i(\x) = \bigg(\int \frac{d^3\k}{(2\pi)^3} W^{i}(\k) \rho_{f}(\k) e^{i\k\cdot\x}\bigg) ^2 \label{pi_computation}
\end{equation}
where $f \in \{m, h\}$ and $W^{i}(\k)$ is a band-pass filter given by:
\begin{equation}
W^i(\k) = \left\{ \begin{array}{cl}
1  & \mbox{ if } k_\mathrm{min}^i < |\k| < k_\mathrm{max}^i  \\ 
0   &  \mathrm{elsewhere} 
\end{array} \right. \label{eq:filter}
\end{equation}
As explained near Eqs.\ (\ref{eq:pmm_loc_def}), (\ref{eq:phh_loc_def}) above, each such $\pi$-field corresponds intuitively to the locally measured small-scale power (in either the matter or halo field) in a certain $k$-range $(k_{\rm min}^i, k_{\rm max}^i)$.

For the 5 matter-derived fields $\pi^m_i$, we use the $k$-bins ($k_{\rm min}^i$, $k_{\rm max}^i$) = \{~(0.5,~1.0), (1.0,~1.5), (1.5,~2.0), (2.0,~2.5), (2.5,~3.0)~\} \mpchinv.
For the 2 halo-derived fields $\pi^h_i$, we use $k$-bins ($k^i_{\rm min}$, $k^i_{\rm max}$) = $\{(0.5, 1.0), (1.0, 1.5)\}$ \mpchinv.

\subsection{Estimating bias and noise from simulations}
\label{ssec:estimating_bias_and_noise}

In this section, we describe our procedure for estimating the parameters $b_\pi$, $\beta_\pi$, and $N_{\pi\pi'}$ from $N$-body simulations.

    The simplest case is the non-Gaussian bias $\beta_\pi = \partial\bar\pi/(\partial\log\sigma_8)$.
    This is conceptually straightforward: we run two ensembles of simulations with different values $\sigma_8^+$, $\sigma_8^-$ of the cosmological parameter $\sigma_8$.
    For each simulation $s$ and choice of $\sigma_8^\pm$, we spatially average $\pi(\x)$ to obtain a per-simulation mean $\bar\pi_s(\sigma_8^\pm)$.
    We then estimate $\beta_\pi = \partial\bar\pi/(\partial\log\sigma_8)$ by numerically differencing:
    \begin{equation}
    \label{eq:piestimator}
    \hat\beta_\pi \equiv \frac{1}{N_{\rm sim}} \sum_{s=1}^{N_{\rm sim}} 
      \frac{\bar\pi_s(\sigma_8^+) - \bar\pi_s(\sigma_8^-)}{\log(\sigma_8^+) - \log(\sigma_8^-)}
    \end{equation}

    In our study, we use \texttt{s8\_m} and \texttt{s8\_p} suite of simulations produced by the Quijote collaboration and described in detail in \ref{sec:simulations} to estimate $\beta_{\pi}$. The \texttt{s8\_m} and \texttt{s8\_p} simulations are run in pairs, with the same random number generator seed and $\sigma_8^-=0.819$ and $\sigma_8^+=0.849$ respectively.
    Thus we can estimate $\beta_\pi$ by finite difference (Eq.\ (\ref{eq:piestimator})) around the fiducial simulation suite with $\sigma_8=0.834$.
    
   Next, we discuss the Gaussian bias $b_\pi$ and the noise $N_{\pi\pi'}$.
  These parameters are defined by the large-scale power spectra:
  \begin{align}
      P_{m\pi}(k) &= b_\pi P_{mm}(k) \nn \\
      P_{\pi\pi'}(k) &= b_\pi b_{\pi'} P_{mm}(k) + N_{\pi\pi'}
      \label{eq:mode_likeihood_pk}
  \end{align}
  We fit for these parameters directly from simulation, using a mode-based likelihood similar to \S{}VII in \cite{giri_2022, giri_2023}.
  We sketch the construction as follows.
  Given $N$ observables $\pi_1,\cdots,\pi_N$, we define the $(N+1)$-component vector:
  \begin{equation}
     \theta(\k) = \left( \begin{array}{c}
       \delta_m(\k) \\ \pi_1(\k) \\ \vdots \\ \pi_N(\k)
       \end{array} \right)
       \label{eq:mcmc_theta}
  \end{equation}
  We also define the $(N+1)$ by $(N+1)$ covariance matrix $C(k)$ by:
  \begin{equation}
  \big\langle \theta(\k) \theta(\k')^\dag \big\rangle = C(k) \, (2\pi)^3 \delta^3(\k-\k')
  \end{equation}
  The matrix elements of $C(k)$ are given by Eq.~(\ref{eq:mode_likeihood_pk}), and depend on the parameters $b_\pi$, $N_{\pi\pi'}$.
  The model likelihood is given by:
\begin{equation}
    \mathcal{L}(\Theta|\mathcal{D}) \propto \prod_{k}
    \frac{1}{\sqrt{\mbox{Det}\, C(k)}}
    \exp(-\frac{{\mathcal D}(\k)^{\dagger} C(k)^{-1} {\mathcal D}(\k)}{2V}) 
    \label{eq:mode_likelihood}
\end{equation}
which we sample using MCMC sampling code \texttt{emcee}\cite{emcee} using conservative priors on model parameters $(b_\pi, N_{\pi\pi'})$. When additionally constraining $\fnl$ in later sections, we use this same likelihood, with the covariance $C(k)$ generalized to include $\fnl$ dependence using Eq. \eqref{eq:key_conjecture1} and Eq. \eqref{eq:key_conjecture2}. 

    

    

\begin{figure*}
    \begin{subfigure}%
        \centering
        \includegraphics[width=0.47\textwidth]{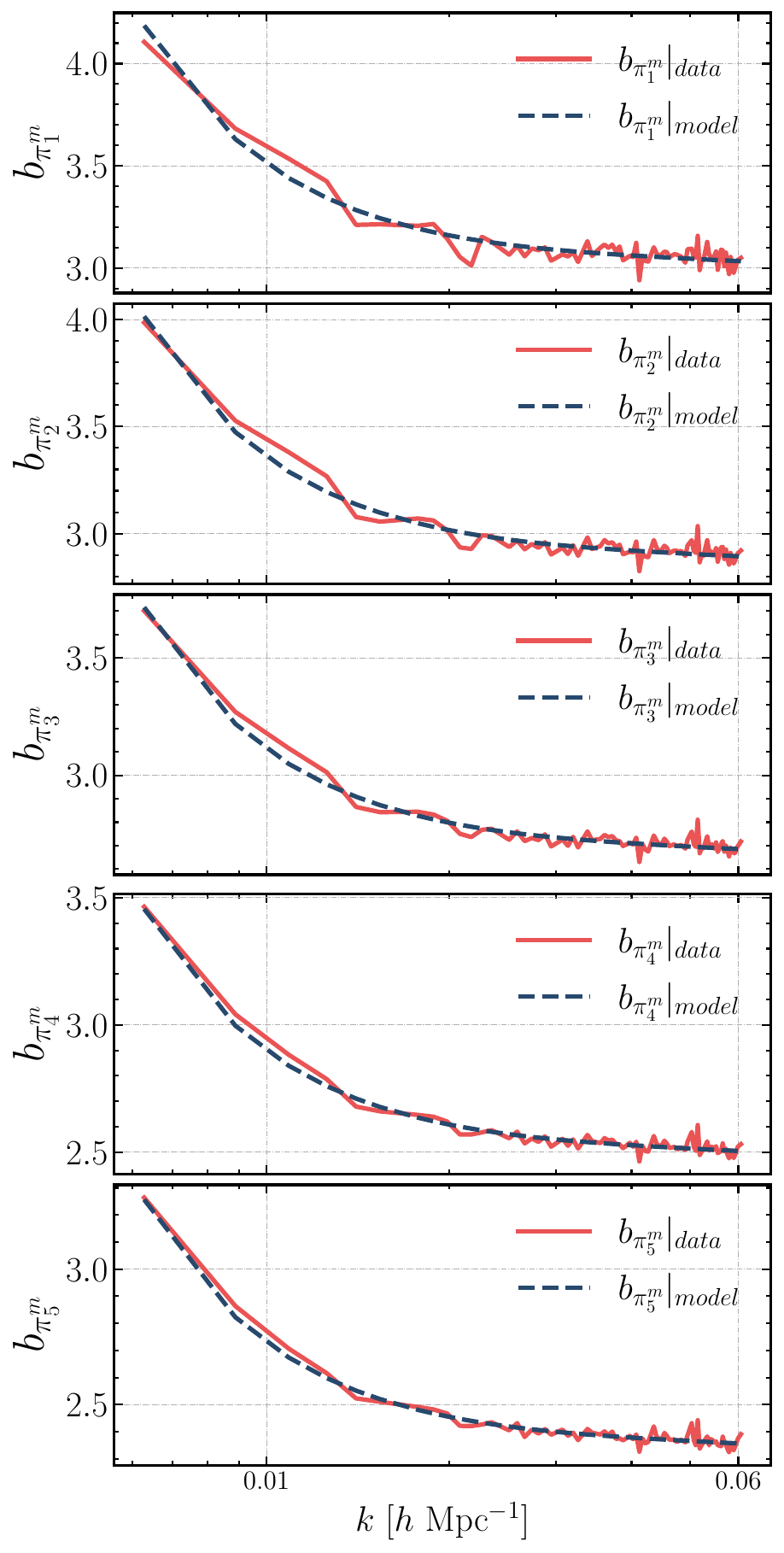}
    \end{subfigure}
    \begin{subfigure}%
        \centering
        \includegraphics[width=0.47\textwidth]{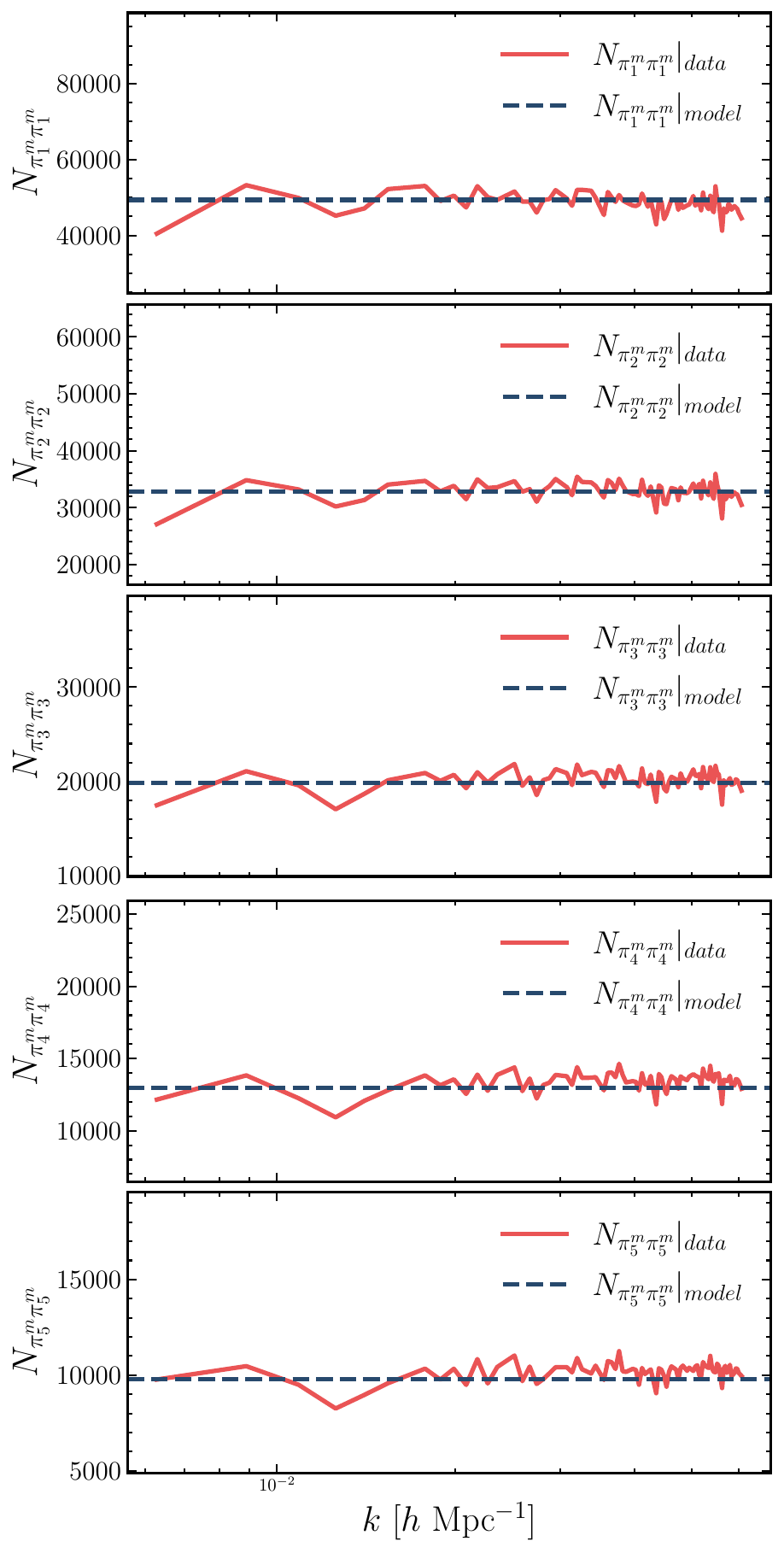}
    \end{subfigure}
    \caption{\emph{Left.} Bias model (\ref{eq:key_conjecture1}) for the five matter-derived $\pi$ fields $\{ \pi^m_i \}_{1\le i\le 5}$ defined in \ref{sec:pi_definition} compared to their empirical bias $b_{\pi^m_i} = P_{m\pi^m_{i}}(k)/P_{mm}(k)$ from simulations, for $\fnl=100$. The best-fit model parameters $(b^G_\pi, N_\pi)$ are obtained from the MCMC pipeline with $k_{\rm max}=0.047$ $h\, \mathrm{Mpc}^{-1}$. The agreement between the simulations and the model is excellent.
    \emph{Right.} Noise power spectra $N_{\pi_i^m \pi_i^m}(k)$ for the $\pi$-fields, computed as described in \S\ref{sec:validation}, showing that the noise power spectrum is constant on large scales.  }
     \label{fig:figure1}
\end{figure*}


\begin{figure*}
    \centering
    \includegraphics[width=1.0\textwidth]{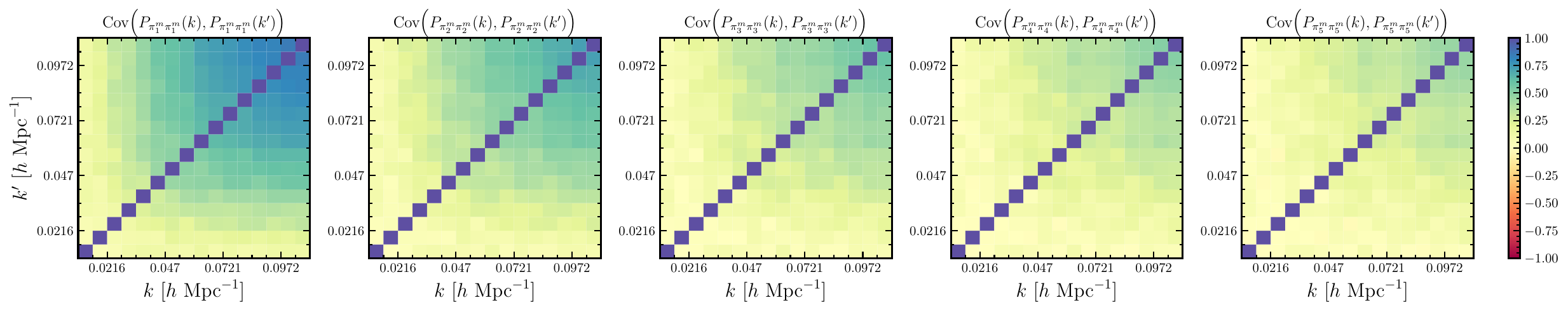}
    \caption{Power spectrum covariance $\mbox{Cov}(P_{\pi^m_i\pi^m_i}(k), P_{\pi^m_i\pi^m_i}(k'))$ for the matter-derived $\pi$-fields. The off-diagonal $(k\ne k')$ covariance is small on large scales. The covariance is estimated using \utkarsh{800} fiducial Quijote simulations. }
    \label{fig:correlation}
\end{figure*}

\begin{figure*}
    \begin{subfigure}%
        \centering
        \includegraphics[width=0.49\textwidth]{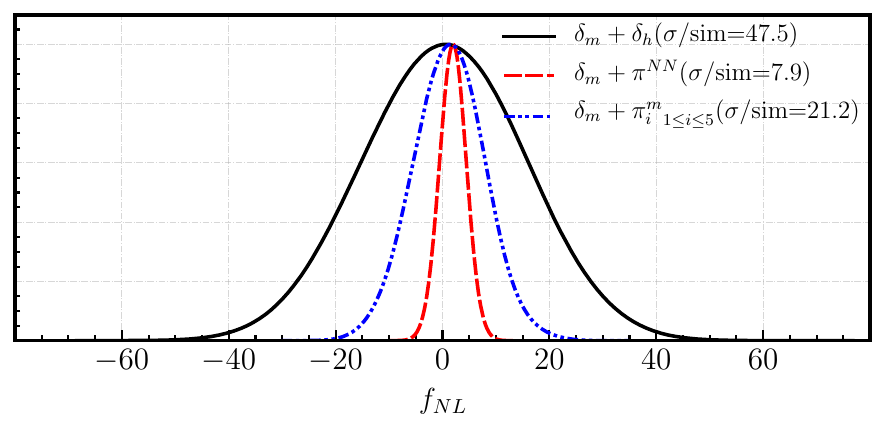}
    \end{subfigure}
    \begin{subfigure}%
        \centering
        \includegraphics[width=0.49\textwidth]{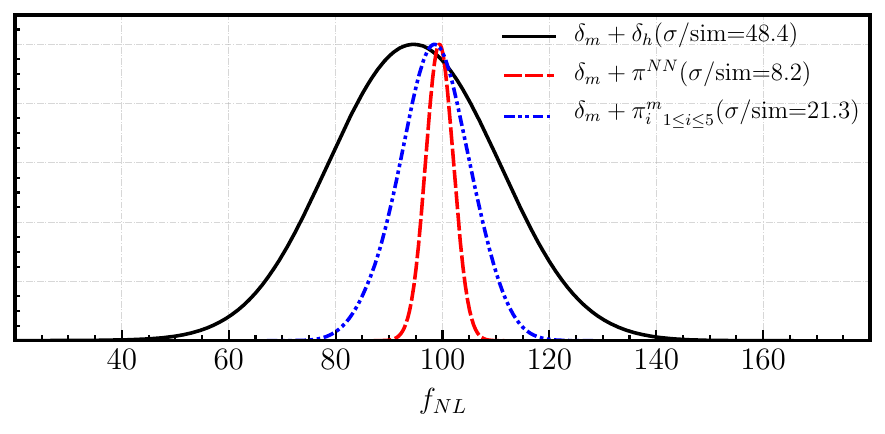}
    \end{subfigure}
    \caption{\emph{Left.} MCMC analysis constraints on $\fnl$ from $\delta_m + \delta_h$ and $\delta_m + \pi^{m}_i$ , using 10 Quijote \texttt{fiducial} simulations with Gaussian initial conditions ($\fnl$=0). The notation $\pi^{m}_i$ refers to the combination of five $\pi^{m}$ fields defined in \ref{sec:pi_definition}.
    These five fields correspond intuitively to locally measured small-scale matter power in five different $k_S$-bins.
    We find that an analysis with the fields $\delta_m + \pi^{m}_{i}$ (blue color) improves error on $\fnl$ by a factor 2.2 compared to a standard $\delta_m + \delta_h$ analysis (black curve). 
    We use modes up to $k_{\rm max}=0.047$ \mpchinv corresponding to the largest 999 modes in the simulation volume. \emph{Right.} Results from analysis of 10 Quijote-PNG \texttt{LC\_p} simulations with $\fnl=100$.  The ``$\sigma$/sim'' value in the label denotes scaled uncertainty on $\fnl$ obtained from a single simulation of volume 1 (\gpchinv)$^3$. {For comparison, we also show the constraints obtained from a neural network based estimate of local power (red color) using a slightly modified version of the architecture presented in \cite{giri_2023} and retraining it using a higher resolution matter field (voxels $=1024^3$) as used in this work to access modes upto $k\sim3$ \mpchinv.  }}
     \label{fig:fnl_constraint}
\end{figure*}

\subsection{Validation on $\fnl$ simulations}
\label{sec:validation}

In this section we analyze \nbody simulations with $\fnl \ne 0$.
Our goal is to validate the key conjectures laid out in \S\ref{ssec:key_conjectures}, which predict clustering observables $P_{m\pi(k)}$ and $P_{\pi\pi'}(k)$ in a non-Gaussian cosmology.
Throughout this section, we take the $\pi$ fields to be the 5 matter-derived fields $\{ \pi^m_i \}_{1\le i\le 5}$ described in \S\ref{sec:pi_definition}.

In the left panels of Fig. \ref{fig:figure1}, we present the empirical bias $b_{\pi^{m}_i} = P_{m\pi^m_{i}}(k)/P_{mm}(k)$ from 50 simulations with $\fnl=100$. The $1/k^2$ behaviour of the bias at the largest-scale is qualitatively evident. We also observe the bias approaching a constant value at $k \sim 0.1$ \mpchinv. For a quantitative comparison, we also show model curves of the form $b(k) = b_\pi + 2 \beta_\pi \fnl / \alpha(k,z)$. Here, the constant bias $b_\pi$ is the best-fit value from an MCMC analysis, but the non-Gaussian bias $\beta_\pi$ is estimated from $\sigma_8^\pm$ simulations using Eq.\ (\ref{eq:piestimator}).

In the right panels of Fig. \ref{fig:figure1}, we present the noise $N_{\pi^m_i \pi^m_i}$ for the $\pi^m_i$ fields. To calculate the noise, we use Gaussian Quijote simulations with $\fnl=0$. The noise is estimated by computing the power spectrum of the residual field defined by $\epsilon_i(\k)=\pi^m_i(k) - b_{\pi^m_i}(k)\delta_m(k)$ where $b_{\pi^m_i}$ is obtained using Eq.~\ref{gaussian_bias_def}. As can be seen, the power spectrum of residual field agrees well with the noise obtained from an MCMC fit. These results conclude our validation of key conjectures 1 and 2 from \S\ref{ssec:key_conjectures}.

Our key conjecture 3 posits that on very large scales, the $\pi$ fields are Gaussian.
As evidence for this conjecture, in Fig.\ \ref{fig:correlation} we show the bandpower covariance $\mbox{Cov}(P_{\pi}(k), P_{\pi}(k'))$ estimated from $N$-body simulations.
For $k \lesssim 0.05$ \mpchinv, the off-diagonals $k\ne k'$ are close to zero, as would be expected for a statistically homogeneous Gaussian field.

\section{Results}
\label{sec:results}

In this section, we consider the central question of this paper: do the squeezed 3-point and collapsed 4-point functions add $\fnl$ information to a matter or halo power spectrum analysis?

In \S\ref{sec:matter_based_analysis}, we will answer this question assuming that the matter field $\delta_m$ can be directly observed on large scales. 
In \ref{halo_based_analysis}, we will assume that only the halo field is observed.

\subsection{Matter field based results from Quijote}
\label{sec:matter_based_analysis}

In our matter-based analysis, we constrain $\fnl$ using large-scale power spectra of the form $P_{m\pi^m_i}(k)$ and $P_{\pi^m_i\pi^m_j}(k)$.
Here, the fields $\{ \pi^{m}_i \}_{1 \le i \le 5}$ were defined in \S\ref{sec:pi_definition}, and correspond to locally measured small-scale matter power in a certain $k$-range $(k_{\rm min}^i, k_{\rm max}^i)$.
In traditional bispectrum language, the observables $P_{m\pi^m_i}(k)$ and $P_{\pi^m_i\pi^m_j}(k)$ correspond respectively to the squeezed matter bispectrum $B_{mmm}$ and collapsed matter trispectrum $T_{mmmm}$.

In the left panel of Fig.~\ref{fig:fnl_constraint}, we present constraints on $\fnl$ from an MCMC analysis which combines $\delta_m$ with the $\pi^{m}_i$ fields, using the mode-based likelihood defined in \S\ref{ssec:estimating_bias_and_noise}. The analysis combines 10 fiducial Quijote simulations with Gaussian initial conditions ($\fnl=0$).
We truncate the sum in Eq.~\ref{eq:mode_likelihood} at a conservative $k_{max}\sim0.047$ \mpchinv corresponding to the largest 999 modes of the simulation volume. 

To describe the MCMC analysis setup in more detail, we denote the six fields $(\delta_m(\k), \pi^m_i(\k))$ by a six-component vector $\theta_i(\k)$ as in Eq.\ (\ref{eq:mcmc_theta}).
The 6$\times$6 covariance $C(k)$ depends on $\fnl$, the Gaussian biases $b_i$, non-Gaussian biases $\beta_i$, and noise parameters $N_{ij}$ (26 parameters total, accounting for the symmetry $N_{ij}=N_{ij}$).
We pre-evaluate the non-Gaussian biases $\beta_i$ by estimating their values in $\sigma_8^\pm$ simulations (see Eq.\ (\ref{eq:piestimator})), and vary the parameters $(\fnl, b_i, N_{ij})$ in the MCMC analysis. {The pre-evaluation is done using 20 pairs of simulations and the estimated biases are ($\beta_h$, $\beta_1$, $\beta_2$, $\beta_3$, $\beta_4$, $\beta_5$) = (0.589, 3.004, 2.924, 2.688, 2.483, 2.347). The statistical scatter in these values is $\sim$1\% }.

In the left panel of Fig.\ \ref{fig:fnl_constraint}, we also show $\fnl$ constraints from a more traditional analysis combining $\delta_m$ and $\delta_h$. We find that the $(\delta_m+\pi^{m})$-based analysis gives a constraint on $\fnl$ which is 2.2 times better than a $\delta_m+\delta_h$ analysis done for the same fiducial simulation volume.
That is, the squeezed matter bispectrum is adding significant $\fnl$ information. 

{Additionally, we also show results obtained from a $(\delta_m+\pi^{NN})$-based analysis where $\pi^{NN}$ is a neural-network based approach for estimating the amplitude of linear fluctuation $\sigma_8$ from a CNN trained on Gaussian \nbody matter field. This CNN based approach was first proposed in \cite{giri_2023} and here, we re-train the architecture on a higher-resolution Quijote matter field voxelized on a grid of size $(1024)^3$ with Nyquist frequency of $\sim \pi$ \mpchinv. Compared to the architecture presented in \cite{giri_2023}, we use a similar architecture with a larger kernel size so that our receptive field is still close  to ~20 \mpchinv corresponding to a $k_{min}\sim 0.3$ \mpchinv. We find that $(\delta_m+\pi^{NN})$-based analysis gives a constraint on $\fnl$ which is 5.5 times better than the $\delta_m+\delta_h$ analysis. As expected, the neural network is even more powerful than the $\pi$ field (with the same robustness, as explained in \cite{giri_2023}), however at the cost of introducing a machine learning element.}

In the right panel of Fig.~\ref{fig:fnl_constraint} we present results from an analysis done using 10 \texttt{LC\_p} simulations from Quijote-PNG simulation suite which have non-Gaussian initial conditions. The \texttt{LC\_p} simulations were run keeping parameters of the simulation similar to the \texttt{fiducial} Quijote run, except for $\fnl$ which was set to 100. %
The true $\fnl$ is within $1\sigma$, and the $\delta_m + \pi$ constraints are again 2.2 times better than the  $\delta_m + \delta_h$ based analysis, consistent with the improvement in the Gaussian case.

\begin{figure}[h]
    \centering
    \includegraphics[width=0.47\textwidth]{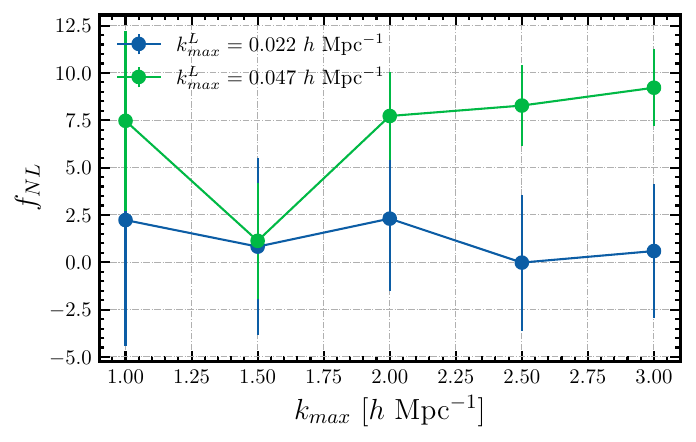}
    \caption{\utkarsh{Mean $\fnl$ estimate from MCMC chains as a function of largest small-scale wavenumber $k_{max}$ included in the analysis.
    We increase $k_{\rm max}$ by adding the fields $\{ \pi^m_i \}_{1\le i\le 5}$ (defined in \S\ref{sec:pi_definition}) one at a time to our MCMC analysis.  The blue and green curves correspond to different choices of the largest large-scale wavenumber $k^L_{max}$.
   The reported mean and error is for 60 independent simulations each with volume of 1 (\gpchinv)$^3$. Large-scale modes were fit using the linear plus $\fnl$ bias model. This shows that adding $\pi$-fields constructed out of very small-scale modes does not bias the $f_{NL}$ as long as our linear bias is fit to large-scale modes of $k^L_{max} \sim 0.02$ \mpchinv}.}
    \label{fig:fnl_bias}
\end{figure}

When we extend our analysis to a larger set of simulations for $\fnl=0$, we find a small ($\Delta\fnl \sim 8$) additive bias, which goes away if we decrease $k_{\rm max}$ from its fiducial value (0.047 \mpchinv). \utkarsh{This is shown in Figure \ref{fig:fnl_bias}).} %
We interpret it as arising from breakdown of the linear bias model ($\pi = b_\pi \delta_m \, + \, \mbox{noise}$) on small scales. For a single simulation volume (1 (\gpchinv)$^3$)  the bias is small. For a larger survey volume, the bias would need to be addressed, either by decreasing $k_{\rm max}$, or by including more terms in the bias model.

So far, we have chosen to use five matter-derived $\pi$-fields $\{ \pi^m_i \}_{1\le i\le 5}$, corresponding to different small-scale $k$-bins (see \S\ref{sec:pi_definition}).
We next explore the impact of varying these choices.

As explained in \ref{sec:formalism}, defining and using multiple $\pi^m$ fields over a $k_s$ window provides more information than using a single $\pi^m$ field over the same $k_s$ window.
Instead of using five $\pi^m_i$ fields covering the range $0.5 \le k_s \le 3.0$, we tried using a single $\pi^m_{\rm coarse}$ field defined over the same $k_s$-range.
We find that the single-simulation $\fnl$ constraint degrades significantly, from $\sigma(\fnl) \sim 25$ to $\sigma(\fnl) \sim 37$.

\begin{figure}[h]
    \centering
    \includegraphics[width=0.45\textwidth]{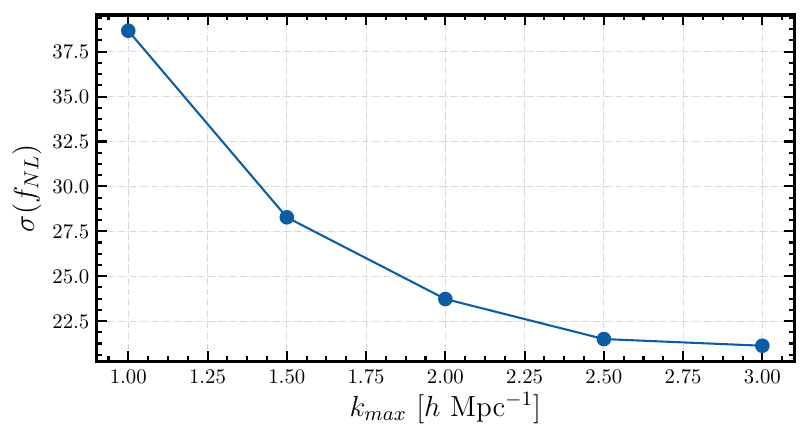}
    \caption{Constraints on $\fnl$ as a function of $k_{\rm max}$, the largest wavenumber where the matter field $\delta_m({\bf k})$ is observed.
    We increase $k_{\rm max}$ by adding the fields $\{ \pi^m_i \}_{1\le i\le 5}$ (defined in \S\ref{sec:pi_definition}) one at a time to our MCMC analysis.
   The reported sensitivity is for a simulation volume of 1 (\gpchinv)$^3$.}
    \label{fig:cumulative_sigma_fnl}
\end{figure}

Finally, to demonstrate how sensitivity to $\fnl$ increases as we go deeper into the non-linear regime, in Fig.~\ref{fig:cumulative_sigma_fnl}, we look at how the constraint $\sigma(\fnl)$ on $\fnl$ evolves as a function of $k$ in our analysis by combining $\delta_m$ with $\pi$ fields which have their k-filter centered at increasingly non-linear scales. As can be seen, the constraints improve monotonically as more and more $\pi$ fields are included in the analysis, suggesting the presence of $\fnl$ sensitive information in the highly non-linear regime. We note that there appears to be almost no extra information contained in the highest $k$ bin, even though Fig. \ref{fig:figure1} shows that the noise in this bin is very low. This means that this bin is highly correlated with the lower $k$ bins. Physically this may be because we have entered the 1-halo regime, but we have not investigated this question in detail.

\begin{figure*}
         \centering
         \includegraphics[width=0.49\textwidth]{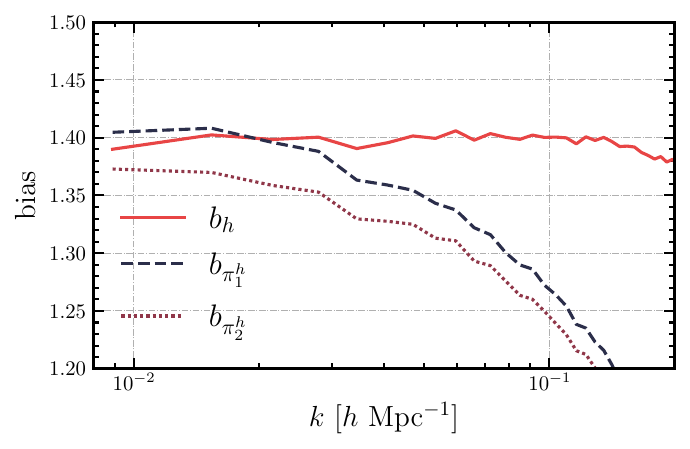}
         \label{fnl_constraint_halo_matter_pi}
     \caption{Bias measured in $\fnl=0$ simulations for the two halo-derived $\pi$ fields $\{ \pi^h_i \}_{1\le i\le 2}$ defined in \ref{sec:pi_definition}, and for the halo field $\delta_h$.
     The curves demonstrate that unlike halo bias $b_h$ which remains constant even at $k=0.1$ \mpchinv, the bias of $\pi^h_i$ starts deviating from a constant at $k \sim 0.015$ \mpchinv. We use this result as a justification for truncating the MCMC likelihood in our halo-based analysis (\S\ref{halo_based_analysis}) at $k_{max}=0.0125$ \mpchinv, corresponding to the 20 largest modes in the simulation volume.}
     \label{fig:halo_based_bias}
\end{figure*}

\begin{figure*}
     \begin{subfigure}%
         \centering
         \includegraphics[width=0.49\textwidth]{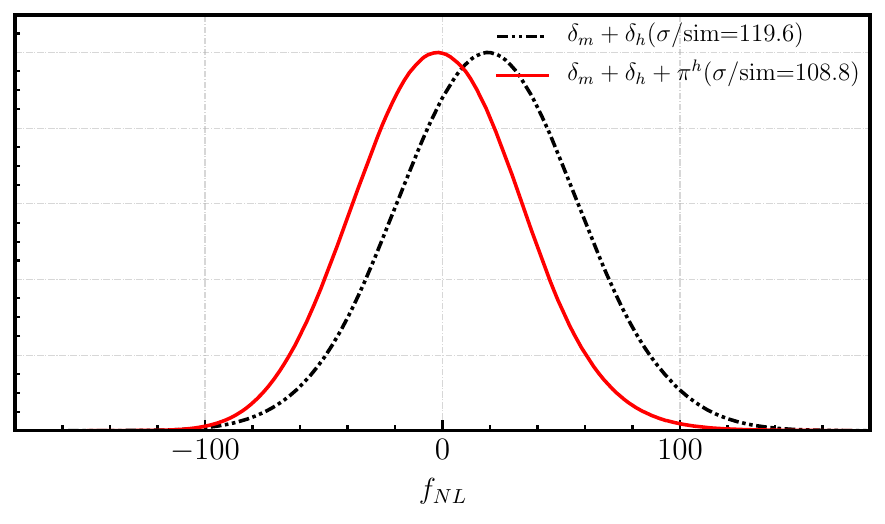}
         \label{fnl_constraint_halo_matter_pi}
     \end{subfigure}
     \begin{subfigure}%
         \centering
         \includegraphics[width=0.49\textwidth]{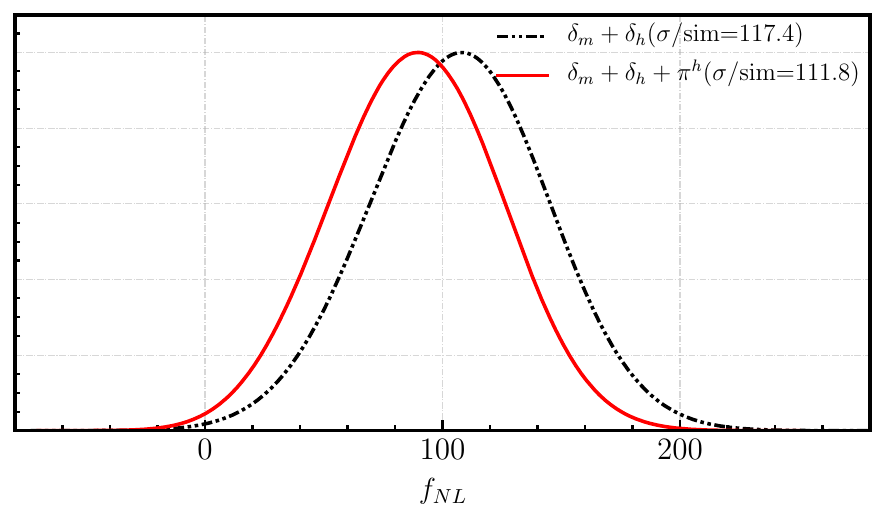}
         \label{fig:fnl_constraint_halo_pi}
     \end{subfigure}
     \caption{\emph{Left.} MCMC analysis constraints on $\fnl$ from $\delta_m + \delta_h$, compared to constraints from $\delta_m + \delta_h + \pi^{h}_i$, where $\pi^{h}_i$ denotes the halo-derived $\pi$-fields $\pi^{h}_1$ and $\pi^h_{2}$ defined in \ref{sec:pi_definition}. The two $\pi^h_i$ fields correspond intuitively to locally measured small-scale halo power in two different $k_S$-bins. \emph{Right.} Constraints on $\fnl$ for  $\fnl=100$ simulations.  %
     In both these cases we use 10 Quijote simulations (same set as that used in Fig.~\ref{fig:fnl_constraint}) with likelihood truncated at $k=0.0125$ $h$ Mpc$^{-1}$ corresponding to the largest 20 modes in the simulation volume. The improvement in $\sigma(\fnl)$ from adding the $\pi^h_i$ fields is marginal.}
     \label{fig:fnl_constraint_halo}
\end{figure*}

\begin{figure}[h]
    \centering
    \includegraphics[width=0.48\textwidth]{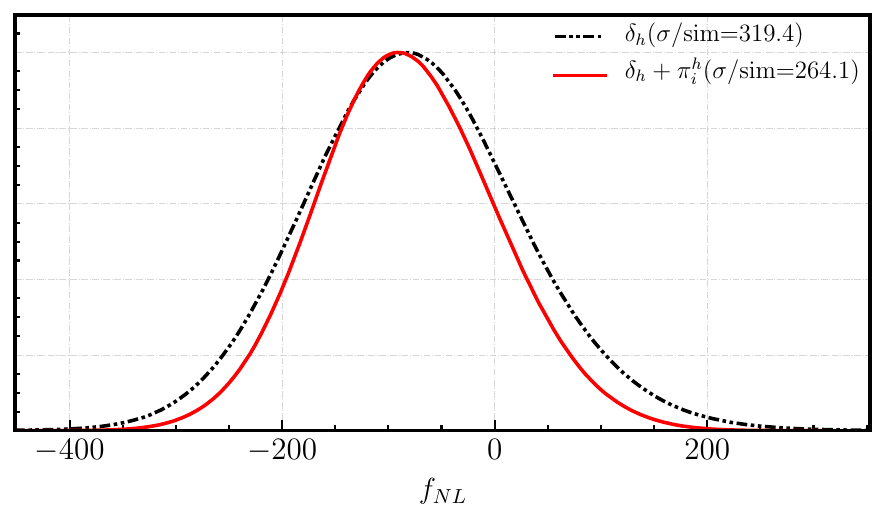}
    \caption{MCMC analysis constraints on $\fnl$ from $\delta_h$ compared to constraints from $\delta_h + \pi^{h}_i$, where $\pi^{h}_i$ denotes the halo-derived $\pi$-fields $\{ \pi^{h}_i \}_{i=1,2}$ defined in \ref{sec:pi_definition}.
    The two $\pi^h_i$ fields correspond intuitively to locally measured small-scale halo power in two different $k_S$-bins. 
    The analysis uses 10 Quijote simulations with Gaussian initial conditions (same set as that used in Fig.~\ref{fig:fnl_constraint}) with likelihood truncated at $k=0.012$ \mpchinv. The noise $N_{\pi_i\pi_j}$ is fixed to the values obtained from $\delta_m + \delta_h + \pi^h$ analysis (Figure \ref{halo_based_analysis}). The improvement in $\sigma(\fnl)$ from adding the $\pi^h_i$ fields is moderate but not negligible (about $25\%$).}
    \label{fig:figure6}
\end{figure}

\subsection{Halo field based results from Quijote}
\label{halo_based_analysis}

In the previous section, we assumed that the matter field was observable on small scales.
In this section, we (more realistically) assume that the small-scale halo field is observed and used to construct the $\pi$ field. We will present two slightly different versions of the analysis, with and without the large-scale matter field, in Figs.\ \ref{fig:fnl_constraint_halo} and \ref{fig:figure6} below.

First we constrain $\fnl$ using large-scale power spectra of the fields $\delta_m$, $\delta_h$, and $\pi^h_i$ (we will drop $\delta_m$ below). Here, the fields  $\{ \pi^h_i \}_{i=1,2}$ were defined in \S\ref{sec:pi_definition}, and correspond to locally measured small-scale halo power in a certain $k$-range $(k_{\rm min}^i, k_{\rm max}^i)$.

In Fig. \ref{fig:halo_based_bias}, we show the large-scale bias of the $\smash{\pi^h_i}$-fields:
\begin{equation}
b^G_{\pi^h_i}(k) = \frac{P_{m\pi^h_i}(k)}{P_{mm}(k)}
\end{equation}
and compare it to halo bias $b^G_h$.
As can be seen, the $\pi^h_i$-bias starts deviating from a scale-independent constant value starting at $k\sim0.015$ \mpchinv. Therefore we conclude that our linear bias for the $\pi^h_i$ fields is valid below this scale, and restrict the MCMC analysis %
to $k_{\rm max}=0.0125$ \mpchinv, corresponding to the largest 20 modes of the simulation volume. 

In the left panel of Fig.~\ref{fig:fnl_constraint_halo}, we present marginalized constraints on $\fnl$ from our analysis of $\delta_m + \delta_h$ and compare it to the constraints obtained from $\delta_m + \delta_h + \pi^{h}_{i}$. 
We find that by adding the $\pi^h$ fields to the MCMC analysis, we obtain negligible improvement in our uncertainty on $\fnl$. In the right panel of Fig.~\ref{fig:fnl_constraint_halo}, we present results of an identical analysis but for simulations with $\fnl=100$, showing that the correct value of $\fnl$ is recovered (within statistical errors).

So far, we have assumed that the matter field is observed on large scales (but not small scales), and the halo field is observed on all scales. 
In Fig.~\ref{fig:figure6}, we eliminate the matter field from the analysis, and continue to assume that the halo field is observed on all scales.
The first curve (labelled $\delta_h$) uses the power spectrum $P_{hh}$ on large scales.
The second curve (labelled $\delta_h + \pi^h_i$) uses large-scale power spectra of the form $P_{hh}$, $P_{h\pi^h_i}$, and $P_{\pi^h_i \pi^h_j}$.
In $N$-point language, this analysis includes the squeezed trispectrum $B_{hhh}$ and collapsed trispectrum $T_{hhhh}$.

Due to the limited number of modes being used, we find that it's not possible to constrain all the bias and noise parameters simultaneously with the $\fnl$ parameter. We therefore decide to fix the noise parameters to their values obtained from the $\delta_m + \delta_h + \pi^{h}_{i}$ analysis described above. After fixing the noise parameters in our MCMC pipeline, we run it to obtain constraints on the bias and $\fnl$ and present the marginalized constraint on $\fnl$. We find that in this analysis without the matter field, the relative improvement in $\sigma(\fnl)$ from adding $\pi^h_i$-fields is non-negligible, but still small. We obtain $\sim \utkarsh{20}$\% improvement in the marginalized error bound. In both  cases, we don't observe any systematic bias in the recovered $\fnl$ estimate.

Summarizing, in this section we compared a ``standard'' analysis of large-scale $\delta_m$ and $\delta_h$ fields to an ``extended'' analysis which also includes $\pi^h_i$-fields.
Our main result is that the improvement in $\sigma(\fnl)$ is marginal.

In this analysis, we have used $k_{\rm max} = 0.0125$ \mpchinv, since the $\pi^h_i$-bias in Fig.\ \ref{fig:halo_based_bias} is only constant on the largest scales.
It is possible that by using a higher order bias model, one could include smaller scales.
It seems unlikely to us that increasing $k_{\rm max}$ would change our results qualitatively, for the following reason.
In Figs.\ \ref{fig:fnl_constraint_halo}, \ref{fig:figure6}, the standard and extended analyses use the same value of $k_{\rm max}$.
If $k_{\rm max}$ is increased consistently in both analyses, then both values of $\sigma(\fnl)$ will decrease, but it seems unlikely that the ratio of $\sigma(\fnl)$ values (or $\fnl$ information per mode) would change. 

\subsection{Comparison with previous work}
\label{ssec:comparison}

In this section, we compare our results with related work (described in the introduction) by Goldstein et. al. \cite{Goldstein:2022hgr} and the Quijote-PNG collaboration \cite{qpng1, jung2022quijotepng, coulton2022quijote, Jung:2022gfa}.

The analysis in \cite{Goldstein:2022hgr} uses the squeezed $(k_L \ll k_S)$ matter bispectrum over the following scales (units $h$/Mpc):
\begin{equation}
0.005 < k_L < 0.06 \hspace{1cm} 0.2 < k_S < 0.6 \label{eq:goldstein_scales}
\end{equation}
and finds statistical error $\sigma(\fnl)=12$ at $z=0$ for simulation volume $(2.4 \mbox{ Gpc}/h)^3$.
Scaling this to the Quijote volume $(1 \mbox{ Gpc}/h)^3$, assuming $\sigma(\fnl) \propto V^{-1/2}$, gives $\sigma(\fnl)=45$.

This can be compared to our $(\delta_m + \pi^m_i)$ analysis from \S\ref{sec:matter_based_analysis}, which gives $\sigma(\fnl)=25$, a factor 1.8 better.
There are two major differences between our analysis and \cite{Goldstein:2022hgr} which may be responsible for the different $\sigma(\fnl)$.
First, we use a different range of scales:
\begin{equation}
0.006 < k_L < 0.047 \hspace{1cm} 0.5 < k_S < 3.0
 \label{eq:squeezed_3pt}
\end{equation}
in the same notation as Eq.\ \ref{eq:goldstein_scales}.
Second, our $(\delta_m + \pi^m_i)$ analysis includes collapsed trispectrum information, in addition to squeezed bispectrum information.

Next, we compare our results to the Quijote-PNG \cite{qpng1,jung2022quijotepng,coulton2022quijote,Jung:2022gfa}
analysis of local-type non-Gaussianity.
(The Quijote-PNG papers also analyze equilateral and orthogonal-type non-Gaussianity, but this is outside the scope of this paper.)
The two analyses are closely related: we study the same high-level questions and use the same simulation parameters.
However, the details are very different as follows.

First, the two analyses use $N$-point functions on different scales.
The Quijote-PNG analysis uses two-point and three-point functions with $k_{\rm max}=0.5$ (units $h$ Mpc$^{-1}$ throughout).
In contrast, we use the two-point function on large scales ($k \lesssim 0.05$), the 3-point function in ``squeezed'' configurations (Eq.\ (\ref{eq:squeezed_3pt})), and the 4-point function in ``collapsed'' configurations
\begin{equation}
|{\bf k}_1 + {\bf k}_2| \lesssim 0.05 \hspace{0.5cm} 
0.5 \lesssim \{ k_1,k_2,k_3,k_4 \} \lesssim 3.0
\label{eq:collapsed_4pt}
\end{equation}
Thus, the two analyses are highly complementary.
For the 2-point function, our analysis contains less information (lower $k_{\rm max}$) than Quijote-PNG.
For the 3-point function, neither analysis is a subset of the other.
Finally, our analysis includes some 4-point information, whereas Quijote-PNG does not use the 4-point function.
(We note that since the two analyses are so complementary, it should be possible to combine them in an analysis to tighten $f_{NL}$ constraints further.)

Second, the Quijote-PNG analysis marginalizes $\Lambda$CDM cosmological parameters (and for analyses including halos, a minimum halo mass $M_{\rm min}$), but does not marginalize a complete set of astrophysical nuisance parameters at $k_{\rm max}=0.05$ (e.g.\ EFT coefficients, higher-order halo biases).

In contrast, we have argued that by marginalizing bias and noise parameters $b_\pi$, $N_{\pi\pi'}$, we automatically marginalize over all nuisance parameters in sight (cosmological or astrophysical).
This is because primordial non-Gaussianity produces signals with a distinctive $1/k^2$ scale dependence.

Finally, the details of the forecasting procedure are quite different (MCMC analysis versus Fisher forecast), and in particular we do not need a large suite of $N$-body simulations in order to estimate covariances and cosmological parameter derivatives.

Despite these differences, our conclusions are very similar to Quijote-PNG.
If the matter field can be observed on small scales, then strong $\fnl$ constraints can be obtained.
Quijote-PNG finds $\sigma(\fnl)\sim35$ at $z=0$\footnote{We read off the value $\sigma(\fnl)\sim35$ from Fig.\ 6 of \cite{qpng1}, right column, curve labelled ``Local'' and ``Marg.\ $\Lambda$CDM params''.}, whereas we find $\sigma(\fnl)\utkarsh{\sim22}$, a factor \utkarsh{>1.5} better.
As with our comparison with \cite{Goldstein:2022hgr} above, the improvement could be due either to the different scales in the two analyses, or to our inclusion of four-point information.

On the other hand, if we only have observations of the small-scale halo field, then improvements from using higher $N$-point functions are marginal (compared to a two-point analysis).
We are surprised by the level of agreement with Quijote-PNG, since the analyses are so different (see above), and so we expect that these conclusions are quite robust.

\section{Conclusion}
\label{sec:discussion}

In this work, we have presented a simulation based approach for constraining local primordial non-Gaussianity parameter $\fnl$ from current and upcoming galaxy survey datasets.
The approach involves computing auxiliary fields $\pi$ using small-scale modes of galaxy or halo fields.
The $\pi$-fields are quadratic in $\delta_g$ or $\delta_h$ and intuitively correspond to locally measured small-scale power.
These fields ``encode'' higher-point information, in the sense that large-scale power spectra ($P_{m\pi}$ or $P_{\pi\pi}$) are equivalent to higher $N$-point functions (squeezed bispectrum or collapsed trispectrum).

We have validated our formalism and developed an end-to-end MCMC pipeline to test the constraining power of the approach when applied to matter and halo fields obtained from \nbody simulations.
The main idea is that on large scales, the $\pi$-fields can be modelled as:
\begin{equation}
\pi(\k) = \left( b_\pi + 2\beta_\pi \frac{f_{NL}}{\alpha(k,z)} \right) \delta_m(\k) + \big(\mbox{Gaussian noise} \big) \nonumber
\label{eq:bias_conclusion}
\end{equation}
This simple-looking statement turns out to have several very interesting consequences.

The Gaussianity of the noise means that we can analyze higher $N$-point functions in a simple way by sampling a Gaussian likelihood function (\ref{eq:mode_likelihood}) for the large-scale modes of the $\delta_m$ and $\pi$ fields.
Although the likelihood is Gaussian, it incorporates (via sample variance of the $\pi$-fields) nontrivial higher-point contributions to the bispectrum and trispectrum covariance.
We have tested Gaussianity of the noise in Fig.\ \ref{fig:correlation}.

The simplicity of the bias model (\ref{eq:bias_conclusion}) means that our approach requires minimal modelling.
In fact, in a Gaussian ($\fnl=0$) cosmology there is no modelling at all -- we simply treat $b_\pi$ and $N_{\pi\pi'}$ as free parameters, to be marginalized in our MCMC sampler.
This procedure automatically marginalizes uncertainty in cosmological and astrophysical nuisance parameters, regardless of the details of these parameters.
This is because the $1/k^2$ scaling in (\ref{eq:bias_conclusion}) in an $\fnl \ne 0$ universe is robust to small-scale peculiarities like aliasing and resolution effects as well as poorly understood baryonic physics. 
We have tested the bias model (\ref{eq:bias_conclusion}) directly in 
Figs.\ \ref{fig:figure1}, \ref{fig:halo_based_bias}.
Additionally, we have done ``end-to-end'' tests of our analysis, by verifying that we recover unbiased $\fnl$ values in simulations (Figs.\ \ref{fig:fnl_constraint}, \ref{fig:fnl_constraint_halo}, \ref{fig:figure6}).

One final advantage of our approach is that it is straightforward to include other large-scale tracer fields.
For example, we could seamlessly include reconstructed kinetic Sunyaev-Zel’dovich velocity \cite{Smith:2018bpn, Munchmeyer:2018eey, giri_2022} for sample variance cancellation. 

\utkarsh{We now comment on some limitations of our study. First, our numerical results are based on the Quijote simulations, which have emerged as a benchmark setup to compare different methods. For the halo resolution of Quijote, we find rather modest gains in sensitivity with our method. Intuitively this is because the shot noise at non-linear scales is large in Quijote, so that only a moderate amount of extra information can be extracted over the number density of halos. However, we expect that the gains from our method would improve in simulations with lower shot noise. Further, while for Quijote a single mass bin analysis is sufficient, a more high-resolution simulation analysis would have to take into account different halo mass bins (or different galaxy samples if available).}

\utkarsh{We do not expect that super sample variance will lead to significant biases, since we do not propose using survey-averaged quantities (such as the mean $\bar\pi$ over the survey volume) as a source of $\fnl$ information. Instead, we use \say{differential} measurements such as $P_{g\pi}(k)$ at nonzero k. In this case, super sample variance can lead to small changes (perhaps a few percent) in our parameters $(b_\pi, \beta_\pi, N_\pi)$. This is not serious since these parameters are either marginalized, or only affect the multiplicative normalization of $\fnl$ (i.e. there is no "additive" bias which can fake an $\fnl$ detection).}

\utkarsh{For simplicity, we also do not apply redshift-space distortions (RSDs) in this work. We conjecture that in the presence of RSDs, our bias model would get a new term
\begin{equation}
\pi(\k) = (b_{\pi} + 2 \beta_{\pi} \fnl / \alpha(k,z) + f \mu^2) \delta_m(\k) + \epsilon
\end{equation}
and that the new term would not significantly bias $f_{NL}$, or increase statistical errors on $\fnl$. (These results are well established for the halo field $\pi=h$, but do not depend on specific properties of the halo field, so we expect that they will also hold for our more general $\pi$ field.) It would be interesting to verify these conjectures in future work.}

\utkarsh{An important aspect of our method is that including very high $k_{max}$ (of the small-scale field) does not lead to biased $\fnl$ estimation. 
This is shown in Fig.\ \ref{fig:fnl_constraint}, where no $\fnl$ bias is seen in the likelihoods labelled ``$\delta_m+\pi^{m_i}$''. These likelihoods include $\pi$ fields constructed from small-scale modes with $k\sim3$ h/Mpc. 
The power of the scale dependent bias formalism is that the parameter combination $\beta_{NL} \ f_{NL}$, and thus the detection significance of non-Gaussianity, cannot be biased by non-linear physics due to the equivalence principle.}

\utkarsh{The scale dependent bias formalism in its current form requires that the halo bias is constant on large scales, which limits the $k_{max}$ of the large-scale fields used in our analysis. One could potentially add a higher order biasing model to go to more non-linear scales, and perhaps increase the $\fnl$ sensitivity somewhat. In future work, we may try to extend the bias model to higher order.}

Remarkably, we find that our simple, low computational cost procedure gives results which are qualitatively consistent with previous studies \cite{qpng1, jung2022quijotepng, coulton2022quijote, Jung:2022gfa, Goldstein:2022hgr}.
In fact, if the matter field can be directly observed on small scales, then our $\sigma(\fnl)$ is a little better than values reported in these studies (see \S\ref{ssec:comparison}).
This could be either because our procedure includes collapsed 4-point information, or because we can use deeply nonlinear modes at very high $k$.
Indeed, because we do not need to model these scales in detail, we can extract $\fnl$ sensitive information from extremely non-linear scales approaching $k_{\rm nyquist}$. 
In practice, we find that as $k_{\rm max}$ is increased, $\sigma(\fnl)$ decreases slowly and eventually saturates (Fig.\ \ref{fig:cumulative_sigma_fnl}).

There is one type of parameter which we do need to model: non-Gaussian biases $\beta_h$, $\beta_\pi$.
This is an issue for essentially all proposals for constraining $\fnl$ from large-scale structure, including non-Gaussian halo bias \cite{Barreira:2022sey, Barreira:2021ueb}, and the discussion below applies generally.
In the simulation-based approach of this paper, we assume perfect knowledge of non-Gaussian biases $\beta_h$ and $\beta_{\pi}$, which we compute following Eq.~\ref{eq:piestimator}.
However, in a more realistic setup, these parameters are not known in advance, and would need to be modelled somehow.
For example, we could use $N$-body simulations (with astrophysical parameters varied over some reasonable range), perturbation theory, or the halo model.
In practice, non-Gaussian biases are degenerate with $\fnl$ (they always appear in the combination $\beta_\pi \fnl$), and so incorrect modelling of $\beta_\pi$ cannot ``fake'' a detection of nonzero $\fnl$ (only the $\fnl$ normalization).
For this reason, we have de-emphasized the issue in this initial study.

When we apply our methods to the halo field $\delta_h$, instead of assuming the matter field $\delta_m$ is directly measurable, we find only marginal improvements (\S\ref{halo_based_analysis}).
The squeezed halo bispectrum $B_{hhh}$ and collapsed halo trispectrum $T_{hhhh}$ add little $\fnl$ information to the halo power spectrum $P_{hh}$.
Similar results were found by the Quijote-PNG collaboration \cite{qpng1, jung2022quijotepng, coulton2022quijote, Jung:2022gfa}
using very different assumptions and methods.
\utkarsh{However, both studies use the Quijote simulations, where the mass resolution is modest and halo number densities are fairly small. In this "1-halo dominated" regime, the $\pi$ fields will be highly correlated with the halo field $\delta_h$ (and with each other) on large scales. (Intuitively, if the global power spectrum is dominated by its 1-halo term, then the locally measured power spectrum $\pi^{loc}$ is highly correlated with the local number density $\delta_h$.) Therefore, It seems plausible to us that higher $N$-point functions may be more useful at higher mass resolution, where halo number densities are larger.}
In a future study, we plan to explore $\fnl$ sensitivity for a higher tracer density sample from simulations like AbacusSummit \cite{maksimova, garrison}.

\section*{Acknowledgements}
We thank William Coulton, Sam Goldstein, Oliver Philcox and Francisco Villaescusa-Navarro for useful comments on the manuscript. \utkarsh{We thank Yurii Kvasiuk for pointing out a missing factor in our likelihood code.}
Part of this work was performed at the Aspen Center for Physics, which is supported by National Science Foundation grant PHY-1607611.
MM acknowledges support from DOE grant DE-SC0022342. Support for this research was provided by the University of Wisconsin - Madison Office of the Vice Chancellor for Research and Graduate Education with funding from the Wisconsin Alumni Research Foundation.
KMS was supported by an NSERC Discovery Grant and a CIFAR fellowship. 
Research at Perimeter Institute is supported in part by the Government of Canada through the Department of Innovation, Science and Economic Development Canada and by the Province of Ontario through the Ministry of Colleges and Universities.
Perimeter Institute's HPC system ``Symmetry'' was used to perform some of the analysis presented in the letter. We have extensively used several python libraries including \textsc{numpy}\cite{harris2020array}, \textsc{matplotlib}\cite{Hunter:2007}, \textsc{CLASS}\cite{Blas_2011}, \textsc{getdist}\cite{Lewis:2019xzd} and \textsc{SciencePlots}\cite{SciencePlots}.

\bibliography{main}

@article{Smith:2007rg,
    author = "Smith, Kendrick M. and Zahn, Oliver and Dore, Olivier",
    title = "{Detection of Gravitational Lensing in the Cosmic Microwave Background}",
    eprint = "0705.3980",
    archivePrefix = "arXiv",
    primaryClass = "astro-ph",
    doi = "10.1103/PhysRevD.76.043510",
    journal = "Phys. Rev. D",
    volume = "76",
    pages = "043510",
    year = "2007"
}

@article{Jeong_2009,
   title={PRIMORDIAL NON-GAUSSIANITY, SCALE-DEPENDENT BIAS, AND THE BISPECTRUM OF GALAXIES},
   volume={703},
   ISSN={1538-4357},
   url={http://dx.doi.org/10.1088/0004-637X/703/2/1230},
   DOI={10.1088/0004-637x/703/2/1230},
   number={2},
   journal={The Astrophysical Journal},
   publisher={American Astronomical Society},
   author={Jeong, Donghui and Komatsu, Eiichiro},
   year={2009},
   eprint={arXiv:0904.0497},
   month={Sep},
   pages={1230–1248}
}

@article{Baldauf_2011,
   title={Primordial non-Gaussianity in the bispectrum of the halo density field},
   volume={2011},
   ISSN={1475-7516},
   url={http://dx.doi.org/10.1088/1475-7516/2011/04/006},
   DOI={10.1088/1475-7516/2011/04/006},
   number={04},
   journal={Journal of Cosmology and Astroparticle Physics},
   publisher={IOP Publishing},
   author={Baldauf, Tobias and Seljak, Uroš and Senatore, Leonardo},
   year={2011},
   eprint={arXiv:1011.1513},
   month={Apr},
   pages={006–006}
}

@article{Schmittfull_2015,
   title={Near optimal bispectrum estimators for large-scale structure},
   volume={91},
   ISSN={1550-2368},
   url={http://dx.doi.org/10.1103/PhysRevD.91.043530},
   DOI={10.1103/physrevd.91.043530},
   number={4},
   journal={Physical Review D},
   publisher={American Physical Society (APS)},
   author={Schmittfull, Marcel and Baldauf, Tobias and Seljak, Uroš},
   year={2015},
   eprint={arXiv:1411.6595},
   month={Feb}
}

@article{Chiang_2017,
   title={Halo squeezed-limit bispectrum with primordial non-Gaussianity: A power spectrum response approach},
   volume={95},
   ISSN={2470-0029},
   url={http://dx.doi.org/10.1103/PhysRevD.95.123517},
   DOI={10.1103/physrevd.95.123517},
   number={12},
   journal={Physical Review D},
   publisher={American Physical Society (APS)},
   author={Chiang, Chi-Ting},
   year={2017},
   month={Jun},
   eprint={arXiv:1701.03374}
}

@misc{deputter2018primordial,
      title={Primordial physics from large-scale structure beyond the power spectrum}, 
      author={Roland de Putter},
      year={2018},
      eprint={arXiv:1802.06762}
}

@article{Dizgah_2020,
   title={Capturing non-Gaussianity of the large-scale structure with weighted skew-spectra},
   volume={2020},
   ISSN={1475-7516},
   url={http://dx.doi.org/10.1088/1475-7516/2020/04/011},
   DOI={10.1088/1475-7516/2020/04/011},
   number={04},
   journal={Journal of Cosmology and Astroparticle Physics},
   publisher={IOP Publishing},
   author={Dizgah, Azadeh Moradinezhad and Lee, Hayden and Schmittfull, Marcel and Dvorkin, Cora},
   year={2020},
   month={Apr},
   pages={011–011},
   eprint={arXiv:1911.05763}
}

@article{Dai_2020,
   title={What can we learn by combining the skew spectrum and the power spectrum?},
   volume={2020},
   ISSN={1475-7516},
   url={http://dx.doi.org/10.1088/1475-7516/2020/08/007},
   DOI={10.1088/1475-7516/2020/08/007},
   number={08},
   journal={Journal of Cosmology and Astroparticle Physics},
   publisher={IOP Publishing},
   author={Dai, Ji-Ping and Verde, Licia and Xia, Jun-Qing},
   year={2020},
   eprint={arXiv:2002.09904},
   month={Aug},
   pages={007–007},
   eprint={arXiv:2002.09904}
}

@misc{darwish2020density,
      title={Density reconstruction from biased tracers and its application to primordial non-Gaussianity}, 
      author={Omar Darwish and Simon Foreman and Muntazir M. Abidi and Tobias Baldauf and Blake D. Sherwin and P. Daniel Meerburg},
      year={2020},
      eprint={arXiv:2007.08472}
}

@article{Cheng__2020,
   title={A new approach to observational cosmology using the scattering transform},
   volume={499},
   ISSN={1365-2966},
   url={http://dx.doi.org/10.1093/mnras/staa3165},
   DOI={10.1093/mnras/staa3165},
   number={4},
   journal={Monthly Notices of the Royal Astronomical Society},
   publisher={Oxford University Press (OUP)},
   author={Cheng, Sihao and Ting, Yuan-Sen and Ménard, Brice and Bruna, Joan},
   year={2020},
   month={Oct},
   pages={5902–5914},
   eprint={arXiv:2006.08561}
}

@article{Villaescusa-Navarro:2020rxg,
    author = "Villaescusa-Navarro, Francisco and others",
    title = "{The CAMELS project: Cosmology and Astrophysics with MachinE Learning Simulations}",
    eprint = "2010.00619",
    archivePrefix = "arXiv",
    primaryClass = "astro-ph.CO",
    doi = "10.3847/1538-4357/abf7ba",
    journal = "Astrophys. J.",
    volume = "915",
    pages = "71",
    year = "2021"
}

@article{Alvarez:2014vva,
    author = "Alvarez, Marcelo and others",
    title = "{Testing Inflation with Large Scale Structure: Connecting Hopes with Reality}",
    eprint = "1412.4671",
    archivePrefix = "arXiv",
    primaryClass = "astro-ph.CO",
    month = "12",
    year = "2014"
}

@article{Villaescusa-Navarro:2019bje,
    author = "Villaescusa-Navarro, Francisco and others",
    title = "{The Quijote simulations}",
    eprint = "1909.05273",
    archivePrefix = "arXiv",
    primaryClass = "astro-ph.CO",
    doi = "10.3847/1538-4365/ab9d82",
    journal = "Astrophys. J. Suppl.",
    volume = "250",
    number = "1",
    pages = "2",
    year = "2020"
}

@article{Hand:2017pqn,
    author = "Hand, Nick and Feng, Yu and Beutler, Florian and Li, Yin and Modi, Chirag and Seljak, Uros and Slepian, Zachary",
    title = "{nbodykit: an open-source, massively parallel toolkit for large-scale structure}",
    eprint = "1712.05834",
    archivePrefix = "arXiv",
    primaryClass = "astro-ph.IM",
    doi = "10.3847/1538-3881/aadae0",
    journal = "Astron. J.",
    volume = "156",
    number = "4",
    pages = "160",
    year = "2018"
}

@article{Slosar_2008,
	doi = {10.1088/1475-7516/2008/08/031},
	url = {https://doi.org/10.1088%2F1475-7516%2F2008%2F08%2F031},
	year = 2008,
	month = {aug},
	publisher = {{IOP} Publishing},
	volume = {2008},
	number = {08},
	pages = {031},
	author = {An{\v{z}
}e Slosar and Christopher Hirata and Uro{\v{s}} Seljak and Shirley Ho and Nikhil Padmanabhan},
	title = {Constraints on local primordial non-Gaussianity from large scale structure},
	journal = {Journal of Cosmology and Astroparticle Physics}
}

@article{DESI:2016fyo,
  title   = {The DESI Experiment Part I: Science,Targeting, and Survey Design},
  author = "Aghamousa, Amir and others",
  year    = {2016},
  journal = {arXiv preprint arXiv: Arxiv-1611.00036}
}

@article{dore2014cosmology,
  title   = {Cosmology with the SPHEREX All-Sky Spectral Survey},
  author  = {Olivier Doré and others},
  year    = {2014},
  journal = {arXiv preprint arXiv: Arxiv-1412.4872}
}

@article{LSSTScience:2009jmu,
    author = "Abell, Paul A. and others",
    collaboration = "LSST Science, LSST Project",
    title = "{LSST Science Book, Version 2.0}",
    eprint = "0912.0201",
    archivePrefix = "arXiv",
    primaryClass = "astro-ph.IM",
    reportNumber = "FERMILAB-TM-2495-A, SLAC-R-1031",
    month = "12",
    year = "2009"
}

@article{Dalal:2007cu,
    author = "Dalal, Neal and Dore, Olivier and Huterer, Dragan and Shirokov, Alexander",
    title = "{The imprints of primordial non-gaussianities on large-scale structure: scale dependent bias and abundance of virialized objects}",
    eprint = "0710.4560",
    archivePrefix = "arXiv",
    primaryClass = "astro-ph",
    doi = "10.1103/PhysRevD.77.123514",
    journal = "Phys. Rev. D",
    volume = "77",
    pages = "123514",
    year = "2008"
}

@article{Biagetti:2020skr,
    author = "Biagetti, Matteo and Cole, Alex and Shiu, Gary",
    title = "{The Persistence of Large Scale Structures I: Primordial non-Gaussianity}",
    eprint = "2009.04819",
    archivePrefix = "arXiv",
    primaryClass = "astro-ph.CO",
    doi = "10.1088/1475-7516/2021/04/061",
    journal = "JCAP",
    volume = "04",
    pages = "061",
    year = "2021"
}

@article{MoradinezhadDizgah:2020whw,
    author = "Moradinezhad Dizgah, Azadeh and Biagetti, Matteo and Sefusatti, Emiliano and Desjacques, Vincent and Nore\~na, Jorge",
    title = "{Primordial Non-Gaussianity from Biased Tracers: Likelihood Analysis of Real-Space Power Spectrum and Bispectrum}",
    eprint = "2010.14523",
    archivePrefix = "arXiv",
    primaryClass = "astro-ph.CO",
    doi = "10.1088/1475-7516/2021/05/015",
    journal = "JCAP",
    volume = "05",
    pages = "015",
    year = "2021"
}

@article{Munchmeyer:2018eey,
    author = {M\"unchmeyer, Moritz and Madhavacheril, Mathew S. and Ferraro, Simone and Johnson, Matthew C. and Smith, Kendrick M.},
    title = "{Constraining local non-Gaussianities with kinetic Sunyaev-Zel\textquoteright{}dovich tomography}",
    eprint = "1810.13424",
    archivePrefix = "arXiv",
    primaryClass = "astro-ph.CO",
    doi = "10.1103/PhysRevD.100.083508",
    journal = "Phys. Rev. D",
    volume = "100",
    number = "8",
    pages = "083508",
    year = "2019"
}

@article{giri_2022,
doi = {10.1088/1475-7516/2022/09/028},
url = {https://dx.doi.org/10.1088/1475-7516/2022/09/028},
year = {2022},
month = {sep},
publisher = {IOP Publishing},
volume = {2022},
number = {09},
pages = {028},
author = {Utkarsh Giri and Kendrick M. Smith},
title = {Exploring KSZ velocity reconstruction with N-body simulations and the halo model},
journal = {Journal of Cosmology and Astroparticle Physics}
}

@article{giri_2023,
  title = {Robust neural network-enhanced estimation of local primordial non-Gaussianity},
  author = {Giri, Utkarsh and M\"unchmeyer, Moritz and Smith, Kendrick M.},
  journal = {Phys. Rev. D},
  volume = {107},
  issue = {6},
  pages = {L061301},
  numpages = {7},
  year = {2023},
  month = {Mar},
  publisher = {American Physical Society},
  doi = {10.1103/PhysRevD.107.L061301},
  url = {https://link.aps.org/doi/10.1103/PhysRevD.107.L061301}
}

@article{Smith:2018bpn,
    author = {Smith, Kendrick M. and Madhavacheril, Mathew S. and M\"unchmeyer, Moritz and Ferraro, Simone and Giri, Utkarsh and Johnson, Matthew C.},
    title = "{KSZ tomography and the bispectrum}",
    eprint = "1810.13423",
    archivePrefix = "arXiv",
    primaryClass = "astro-ph.CO",
    month = "10",
    year = "2018"
}

@article{emcee,
   author = {{Foreman-Mackey}, D. and {Hogg}, D.~W. and {Lang}, D. and {Goodman}, J.},
    title = {emcee: The MCMC Hammer},
  journal = {PASP},
     year = 2013,
   volume = 125,
    pages = {306-312},
   eprint = {1202.3665},
      doi = {10.1086/670067}
}

@article{Biagetti_2017,
	doi = {10.1093/mnras/stx714},
  
	url = {https://doi.org/10.1093%2Fmnras%2Fstx714},
  
	year = 2017,
	month = {mar},
  
	publisher = {Oxford University Press ({OUP})},
  
	volume = {468},
  
	number = {3},
  
	pages = {3277--3288},
  
	author = {Matteo Biagetti and Titouan Lazeyras and Tobias Baldauf and Vincent Desjacques and Fabian Schmidt},
  
	title = {Verifying the consistency relation for the scale-dependent bias from local primordial non-Gaussianity},
  
	journal = {Monthly Notices of the Royal Astronomical Society}
}

@Article{         harris2020array,
 title         = {Array programming with {NumPy}},
 author        = {Charles R. Harris and others},
 year          = {2020},
 month         = sep,
 journal       = {Nature},
 volume        = {585},
 number        = {7825},
 pages         = {357--362},
 doi           = {10.1038/s41586-020-2649-2},
 publisher     = {Springer Science and Business Media {LLC}},
 url           = {https://doi.org/10.1038/s41586-020-2649-2}
}

@article{Blas_2011,
	doi = {10.1088/1475-7516/2011/07/034},
  
	url = {https://doi.org/10.1088%2F1475-7516%2F2011%2F07%2F034},
  
	year = 2011,
	month = {jul},
  
	publisher = {{IOP} Publishing},
  
	volume = {2011},
  
	number = {07},
  
	pages = {034--034},
  
	author = {Diego Blas and Julien Lesgourgues and Thomas Tram},
  
	title = {The Cosmic Linear Anisotropy Solving System ({CLASS}). Part {II}: Approximation schemes},
  
	journal = {Journal of Cosmology and Astroparticle Physics}
}

@Article{Hunter:2007,
  Author    = {Hunter, J. D.},
  Title     = {Matplotlib: A 2D graphics environment},
  Journal   = {Computing in Science \& Engineering},
  Volume    = {9},
  Number    = {3},
  Pages     = {90--95},
  publisher = {IEEE COMPUTER SOC},
  doi       = {10.1109/MCSE.2007.55},
  year      = 2007
}

@article{Lewis:2019xzd,
 author         = "Lewis, Antony",
 title          = "{GetDist: a Python package for analysing Monte Carlo
                   samples}",
 year           = "2019",
 eprint         = "1910.13970",
 archivePrefix  = "arXiv",
 primaryClass   = "astro-ph.IM",
 SLACcitation   = "%%CITATION = ARXIV:1910.13970;%%",
 url            = "https://getdist.readthedocs.io"
}

@article{SciencePlots,
  author       = {John D. Garrett},
  title        = {{garrettj403/SciencePlots}},
  month        = sep,
  year         = 2021,
  publisher    = {Zenodo},
  version      = {1.0.9},
  doi          = {10.5281/zenodo.4106649},
  url          = {http://doi.org/10.5281/zenodo.4106649}
}

@article{Planck:2018vyg,
    author = "Aghanim, N. and others",
    collaboration = "Planck",
    title = "{Planck 2018 results. VI. Cosmological parameters}",
    eprint = "1807.06209",
    archivePrefix = "arXiv",
    primaryClass = "astro-ph.CO",
    doi = "10.1051/0004-6361/201833910",
    journal = "Astron. Astrophys.",
    volume = "641",
    pages = "A6",
    year = "2020",
    note = "[Erratum: Astron.Astrophys. 652, C4 (2021)]"
}

@article{Biagetti_2019,
	doi = {10.3390/galaxies7030071},
  
	url = {https://doi.org/10.3390%2Fgalaxies7030071},
  
	year = 2019,
	month = {aug},
  
	publisher = {{MDPI} {AG}
},
  
	volume = {7},
  
	number = {3},
  
	pages = {71},
  
	author = {Matteo Biagetti},
  
	title = {The Hunt for Primordial Interactions in the Large-Scale Structures of the Universe},
  
	journal = {Galaxies}
}

@article{qpng1,
    author = "Coulton, William R. and Villaescusa-Navarro, Francisco and Jamieson, Drew and Baldi, Marco and Jung, Gabriel and Karagiannis, Dionysios and Liguori, Michele and Verde, Licia and Wandelt, Benjamin D.",
    title = "{Quijote-PNG: Simulations of Primordial Non-Gaussianity and the Information Content of the Matter Field Power Spectrum and Bispectrum}",
    eprint = "2206.01619",
    archivePrefix = "arXiv",
    primaryClass = "astro-ph.CO",
    doi = "10.3847/1538-4357/aca8a7",
    journal = "Astrophys. J.",
    volume = "943",
    number = "1",
    pages = "64",
    year = "2023"
}

@article{coulton2022quijote,
    author = "Coulton, William R. and Villaescusa-Navarro, Francisco and Jamieson, Drew and Baldi, Marco and Jung, Gabriel and Karagiannis, Dionysios and Liguori, Michele and Verde, Licia and Wandelt, Benjamin D.",
    title = "{Quijote-PNG: The Information Content of the Halo Power Spectrum and Bispectrum}",
    eprint = "2206.15450",
    archivePrefix = "arXiv",
    primaryClass = "astro-ph.CO",
    doi = "10.3847/1538-4357/aca7c1",
    journal = "Astrophys. J.",
    volume = "943",
    number = "2",
    pages = "178",
    year = "2023"
}

@article{jung2022quijotepng,
    author = "Jung, Gabriel and Karagiannis, Dionysios and Liguori, Michele and Baldi, Marco and Coulton, William R. and Jamieson, Drew and Verde, Licia and Villaescusa-Navarro, Francisco and Wandelt, Benjamin D.",
    title = "{Quijote-PNG: Quasi-maximum Likelihood Estimation of Primordial Non-Gaussianity in the Nonlinear Dark Matter Density Field}",
    eprint = "2206.01624",
    archivePrefix = "arXiv",
    primaryClass = "astro-ph.CO",
    doi = "10.3847/1538-4357/ac9837",
    journal = "Astrophys. J.",
    volume = "940",
    number = "1",
    pages = "71",
    year = "2022"
}

@article{Barreira:2022sey,
    author = "Barreira, Alexandre",
    title = "{Can we actually constrain f$_{NL}$ using the scale-dependent bias effect? An illustration of the impact of galaxy bias uncertainties using the BOSS DR12 galaxy power spectrum}",
    eprint = "2205.05673",
    archivePrefix = "arXiv",
    primaryClass = "astro-ph.CO",
    doi = "10.1088/1475-7516/2022/11/013",
    journal = "JCAP",
    volume = "11",
    pages = "013",
    year = "2022"
}

@article{Barreira:2021ueb,
    author = "Barreira, Alexandre",
    title = "{Predictions for local PNG bias in the galaxy power spectrum and bispectrum and the consequences for f $_{NL}$ constraints}",
    eprint = "2107.06887",
    archivePrefix = "arXiv",
    primaryClass = "astro-ph.CO",
    doi = "10.1088/1475-7516/2022/01/033",
    journal = "JCAP",
    volume = "01",
    number = "01",
    pages = "033",
    year = "2022"
}

@article{EUCLID:2011zbd,
  title   = {Euclid Definition Study Report},
  author  = {R. Laureijs and others},
  year    = {2011},
  journal = {arXiv preprint arXiv: Arxiv-1110.3193}
}

@article{Springel:2005mi,
    author = "Springel, Volker",
    title = "{The Cosmological simulation code GADGET-2}",
    eprint = "astro-ph/0505010",
    archivePrefix = "arXiv",
    doi = "10.1111/j.1365-2966.2005.09655.x",
    journal = "Mon. Not. Roy. Astron. Soc.",
    volume = "364",
    pages = "1105--1134",
    year = "2005"
}

@MISC{Pylians,
    author = {{Villaescusa-Navarro}, Francisco},
    title = "{Pylians: Python libraries for the analysis of numerical simulations}",
    howpublished = {Astrophysics Source Code Library, record ascl:1811.008},
    year = 2018,
    month = nov,
    eid = {ascl:1811.008},
    pages = {ascl:1811.008},
    archivePrefix = {ascl},
    eprint = {1811.008},
    adsurl = {https://ui.adsabs.harvard.edu/abs/2018ascl.soft11008V}
}

@article{Cabass:2022ymb,
    author = "Cabass, Giovanni and Ivanov, Mikhail M. and Philcox, Oliver H. E. and Simonovi\'c, Marko and Zaldarriaga, Matias",
    title = "{Constraints on multifield inflation from the BOSS galaxy survey}",
    eprint = "2204.01781",
    archivePrefix = "arXiv",
    primaryClass = "astro-ph.CO",
    reportNumber = "CERN-TH-2022-055",
    doi = "10.1103/PhysRevD.106.043506",
    journal = "Phys. Rev. D",
    volume = "106",
    number = "4",
    pages = "043506",
    year = "2022"
}

@article{Jung:2022gfa,
  title   = {Quijote-PNG: Quasi-maximum likelihood estimation of Primordial Non-Gaussianity in the non-linear halo density field},
  author  = {Gabriel Jung and Dionysios Karagiannis and Michele Liguori and Marco Baldi and William R Coulton and Drew Jamieson and Licia Verde and Francisco Villaescusa-Navarro and Benjamin D. Wandelt},
  year    = {2022},
  journal = {arXiv preprint arXiv: Arxiv-2211.07565}
}

@article{Andrews:2022nvv,
    author = "Andrews, Adam and Jasche, Jens and Lavaux, Guilhem and Schmidt, Fabian",
    title = "{Bayesian field-level inference of primordial non-Gaussianity using next-generation galaxy surveys}",
    eprint = "2203.08838",
    archivePrefix = "arXiv",
    primaryClass = "astro-ph.CO",
    doi = "10.1093/mnras/stad432",
    journal = "Mon. Not. Roy. Astron. Soc.",
    volume = "520",
    number = "4",
    pages = "5746--5763",
    year = "2023"
}

@article{chen2010primordial,
  title   = {Primordial Non-Gaussianities from Inflation Models},
  author  = {Xingang Chen},
  year    = {2010},
  journal = {arXiv preprint arXiv: Arxiv-1002.1416}
}

@article{Achucarro:2022qrl,
    author = "Ach\'ucarro, Ana and others",
    title = "{Inflation: Theory and Observations}",
    eprint = "2203.08128",
    archivePrefix = "arXiv",
    primaryClass = "astro-ph.CO",
    month = "3",
    year = "2022"
}

@ARTICLE{maksimova,
    author = {Maksimova, Nina A and Garrison, Lehman H and Eisenstein, Daniel J and Hadzhiyska, Boryana and Bose, Sownak and Satterthwaite, Thomas P},
    title = "{AbacusSummit: a massive set of high-accuracy, high-resolution N-body simulations}",
    journal = {Monthly Notices of the Royal Astronomical Society},
    volume = {508},
    number = {3},
    pages = {4017-4037},
    year = {2021},
    month = {09},
    issn = {0035-8711},
    doi = {10.1093/mnras/stab2484},
    url = {https://doi.org/10.1093/mnras/stab2484},
    eprint = {https://academic.oup.com/mnras/article-pdf/508/3/4017/40811763/stab2484.pdf},
}

@ARTICLE{garrison,
    author = {Garrison, Lehman H and Eisenstein, Daniel J and Ferrer, Douglas and Maksimova, Nina A and Pinto, Philip A},
    title = "{The abacus cosmological N-body code}",
    journal = {Monthly Notices of the Royal Astronomical Society},
    volume = {508},
    number = {1},
    pages = {575-596},
    year = {2021},
    month = {09},
    issn = {0035-8711},
    doi = {10.1093/mnras/stab2482},
    url = {https://doi.org/10.1093/mnras/stab2482},
    eprint = {https://academic.oup.com/mnras/article-pdf/508/1/575/40458823/stab2482.pdf},
}

@article{Angulo_2021,
    author = "Angulo, Raul E. and Hahn, Oliver",
    title = "{Large-scale dark matter simulations}",
    eprint = "2112.05165",
    archivePrefix = "arXiv",
    primaryClass = "astro-ph.CO",
    doi = "10.1007/s41115-021-00013-z",
    month = "12",
    year = "2021"
}

@article{chisari2019modelling,
  title     = {Modelling baryonic feedback for survey cosmology},
  author    = {N. E. Chisari and A. Mead and S. Joudaki and P. Ferreira and A. Schneider and J. Mohr and T. Tröster and D. Alonso and I. McCarthy and S. Martin-Alvarez and J. Devriendt and A. Slyz and Marcel van Daalen},
  journal   = {The Open Journal Of Astrophysics},
  year      = {2019},
  doi       = {10.21105/astro.1905.06082},
  bibSource = {Semantic Scholar https://www.semanticscholar.org/paper/7f17014a24b8d7ba8201992bd0ec9d8b555f6f7e}
}

@article{DAmico:2022gki,
  author        = {D'Amico, Guido and Lewandowski, Matthew and Senatore, Leonardo and Zhang, Pierre},
  title         = {{Limits on primordial non-Gaussianities from BOSS galaxy-clustering data}},
  eprint        = {2201.11518},
  archivePrefix = {arXiv},
  primaryClass  = {astro-ph.CO},
  month         = {1},
  year          = {2022}
}

@ARTICLE{fof,
       author = {{Davis}, M. and {Efstathiou}, G. and {Frenk}, C.~S. and {White}, S.~D.~M.},
        title = "{The evolution of large-scale structure in a universe dominated by cold dark matter}",
      journal = {\apj},
         year = 1985,
        month = may,
       volume = {292},
        pages = {371-394},
          doi = {10.1086/163168},
       adsurl = {https://ui.adsabs.harvard.edu/abs/1985ApJ...292..371D}
}

@BOOK{hockneyeastwood,
       author = {{Hockney}, R.~W. and {Eastwood}, J.~W.},
        title = "{Computer simulation using particles}",
         year = 1988,
       adsurl = {https://ui.adsabs.harvard.edu/abs/1988csup.book.....H}
}

@article{Sullivan:2023qjr,
    author = "Sullivan, James M. and Prijon, Tijan and Seljak, Uros",
    title = "{Learning to Concentrate: Multi-tracer Forecasts on Local Primordial Non-Gaussianity with Machine-Learned Bias}",
    eprint = "2303.08901",
    archivePrefix = "arXiv",
    primaryClass = "astro-ph.CO",
    month = "3",
    year = "2023"
}

@article{Goldstein:2022hgr,
    author = "Goldstein, Samuel and Esposito, Angelo and Philcox, Oliver H. E. and Hui, Lam and Hill, J. Colin and Scoccimarro, Roman and Abitbol, Maximilian H.",
    title = "{Squeezing fNL out of the matter bispectrum with consistency relations}",
    eprint = "2209.06228",
    archivePrefix = "arXiv",
    primaryClass = "astro-ph.CO",
    doi = "10.1103/PhysRevD.106.123525",
    journal = "Phys. Rev. D",
    volume = "106",
    number = "12",
    pages = "123525",
    year = "2022"
}

@article{Desjacques_2008,
    author = "Desjacques, Vincent and Seljak, Uros and Iliev, Ilian",
    title = "{Scale-dependent bias induced by local non-Gaussianity: A comparison to N-body simulations}",
    eprint = "0811.2748",
    archivePrefix = "arXiv",
    primaryClass = "astro-ph",
    doi = "10.1111/j.1365-2966.2009.14721.x",
    journal = "Mon. Not. Roy. Astron. Soc.",
    volume = "396",
    pages = "85--96",
    year = "2009"
}

@article{Crocce_2006,
    author = "Crocce, M. and Pueblas, S. and Scoccimarro, R.",
    title = "{Transients from Initial Conditions in Cosmological Simulations}",
    eprint = "astro-ph/0606505",
    archivePrefix = "arXiv",
    doi = "10.1111/j.1365-2966.2006.11040.x",
    journal = "Mon. Not. Roy. Astron. Soc.",
    volume = "373",
    pages = "369--381",
    year = "2006"
}

@article{Chiang:2014oga,
    author = "Chiang, Chi-Ting and Wagner, Christian and Schmidt, Fabian and Komatsu, Eiichiro",
    title = "{Position-dependent power spectrum of the large-scale structure: a novel method to measure the squeezed-limit bispectrum}",
    eprint = "1403.3411",
    archivePrefix = "arXiv",
    primaryClass = "astro-ph.CO",
    doi = "10.1088/1475-7516/2014/05/048",
    journal = "JCAP",
    volume = "05",
    pages = "048",
    year = "2014"
}

@article{Barreira:2019icq,
    author = "Barreira, Alexandre",
    title = "{The squeezed matter bispectrum covariance with responses}",
    eprint = "1901.01243",
    archivePrefix = "arXiv",
    primaryClass = "astro-ph.CO",
    doi = "10.1088/1475-7516/2019/03/008",
    journal = "JCAP",
    volume = "03",
    pages = "008",
    year = "2019"
}

@article{Sailer:2021yzm,
    author = "Sailer, Noah and Castorina, Emanuele and Ferraro, Simone and White, Martin",
    title = "{Cosmology at high redshift \textemdash{} a probe of fundamental physics}",
    eprint = "2106.09713",
    archivePrefix = "arXiv",
    primaryClass = "astro-ph.CO",
    doi = "10.1088/1475-7516/2021/12/049",
    journal = "JCAP",
    volume = "12",
    number = "12",
    pages = "049",
    year = "2021"
}

@article{Floss:2022wkq,
    author = {Fl\"oss, Thomas and Biagetti, Matteo and Meerburg, P. Daniel},
    title = "{Primordial non-Gaussianity and non-Gaussian covariance}",
    eprint = "2206.10458",
    archivePrefix = "arXiv",
    primaryClass = "astro-ph.CO",
    doi = "10.1103/PhysRevD.107.023528",
    journal = "Phys. Rev. D",
    volume = "107",
    number = "2",
    pages = "023528",
    year = "2023"
}

@article{Peloso,
    author = "Peloso, Marco and Pietroni, Massimo",
    title = "{Galilean invariance and the consistency relation for the nonlinear squeezed bispectrum of large scale structure}",
    eprint = "1302.0223",
    archivePrefix = "arXiv",
    primaryClass = "astro-ph.CO",
    reportNumber = "UMN-TH-3134-13",
    doi = "10.1088/1475-7516/2013/05/031",
    journal = "JCAP",
    volume = "05",
    pages = "031",
    year = "2013"
}

@article{Kehagias,
    author = "Kehagias, A. and Riotto, A.",
    title = "{Symmetries and Consistency Relations in the Large Scale Structure of the Universe}",
    eprint = "1302.0130",
    archivePrefix = "arXiv",
    primaryClass = "astro-ph.CO",
    doi = "10.1016/j.nuclphysb.2013.05.009",
    journal = "Nucl. Phys. B",
    volume = "873",
    pages = "514--529",
    year = "2013"
}

@phdthesis{Simonovic,
    author = "Simonovic, Marko",
    title = "{Cosmological Consistency Relations}",
    school = "SISSA, Trieste",
    year = "2014"
}

@article{Friedrich2019,
    author = "Friedrich, Oliver and Uhlemann, Cora and Villaescusa-Navarro, Francisco and Baldauf, Tobias and Manera, Marc and Nishimichi, Takahiro",
    title = "{Primordial non-Gaussianity without tails \textendash{} how to measure fNL with the bulk of the density PDF}",
    eprint = "1912.06621",
    archivePrefix = "arXiv",
    primaryClass = "astro-ph.CO",
    reportNumber = "YITP-19-122",
    doi = "10.1093/mnras/staa2160",
    journal = "Mon. Not. Roy. Astron. Soc.",
    volume = "498",
    number = "1",
    pages = "464--483",
    year = "2020"
}

@article{Komatsu:2001rj,
    author = "Komatsu, Eiichiro and Spergel, David N.",
    title = "{Acoustic signatures in the primary microwave background bispectrum}",
    eprint = "astro-ph/0005036",
    archivePrefix = "arXiv",
    doi = "10.1103/PhysRevD.63.063002",
    journal = "Phys. Rev. D",
    volume = "63",
    pages = "063002",
    year = "2001"
}

@article{Esposito:2019jkb,
    author = "Esposito, Angelo and Hui, Lam and Scoccimarro, Roman",
    title = "{Nonperturbative test of consistency relations and their violation}",
    eprint = "1905.11423",
    archivePrefix = "arXiv",
    primaryClass = "astro-ph.CO",
    doi = "10.1103/PhysRevD.100.043536",
    journal = "Phys. Rev. D",
    volume = "100",
    number = "4",
    pages = "043536",
    year = "2019"
}

@article{dePutter:2016moa,
    author = "de Putter, Roland and Dor\'e, Olivier and Green, Daniel and Meyers, Joel",
    title = "{Single-Field Inflation and the Local Ansatz: Distinguishability and Consistency}",
    eprint = "1610.00785",
    archivePrefix = "arXiv",
    primaryClass = "hep-th",
    doi = "10.1103/PhysRevD.95.063501",
    journal = "Phys. Rev. D",
    volume = "95",
    number = "6",
    pages = "063501",
    year = "2017"
}

@article{Smith:2011if,
  author        = {Smith, Kendrick M. and LoVerde, Marilena and Zaldarriaga, Matias},
  title         = {A universal bound on N-point correlations from inflation},
  eprint        = {1108.1805},
  archivePrefix = {arXiv},
  primaryClass  = {astro-ph.CO},
  doi           = {10.1103/PhysRevLett.107.191301},
  journal       = {Phys. Rev. Lett.},
  volume        = {107},
  pages         = {191301},
  year          = {2011}
}

\end{document}